\documentclass[fleqn,usenatbib]{mnras}

\usepackage{newtxtext,newtxmath}

\usepackage[T1]{fontenc}
\usepackage{multirow}

\DeclareRobustCommand{\VAN}[3]{#2}
\let\VANthebibliography\thebibliography
\def\thebibliography{\DeclareRobustCommand{\VAN}[3]{##3}\VANthebibliography}

\usepackage{graphicx}	
\usepackage{amsmath}	
\usepackage{xspace}     
\usepackage{float}      

\newcommand{\nicer}{\textit{NICER}}

\newcommand{\maxi}{\textit{MAXI}}

\newcommand{\gx}{GX339--4}
\newcommand{\dps}{dynamical power-spectrum\xspace}

\title[QPOs during state transitions]{Flip-flop QPO changes during state transitions: a case study of GX339--4 and theoretical discussion}

\author[D. J. K. Buisson et al.]
{D. J. K. Buisson$^{1}$,\thanks{Email: djkbuisson@gmail.com, djkb2@cantab.ac.uk}
G. Marcel$^{2,3,4}$,
V. L\'opez-Barquero$^{4,5}$,
S. E. Motta$^{6}$,
S. G. D. Turner$^{4,7}$,\newauthor
F. M. Vincentelli$^{8}$
\\
$^{1}$Independent Researcher\\
$^{2}$Department of Physics and Astronomy, FI-20014 University of Turku, Finland\\
$^{3}$Nicolaus Copernicus Astronomical Center, Polish Academy of Sciences, Bartycka 18, PL-00-716 Warszawa, Poland\\
$^{4}$Institute of Astronomy, University of Cambridge, Madingley Road, Cambridge CB3 0HA, UK\\
$^{5}$Department of Astronomy, University of Maryland, College Park, MD 20742-2421, USA\\
$^{6}$Istituto Nazionale di Astroﬁsica, Osservatorio Astronomico di Brera, via E. Bianchi 46, 23807 Merate, (LC), Italy\\
$^{7}$Department of Applied Mathematics and Theoretical Physics, University of Cambridge, Wilberforce Road, Cambridge CB3 0WA, UK\\
$^{8}$School of Physics and Astronomy, University of Southampton, Highfield, Southampton, SO17 1BJ, UK\\
}

\date{11\textsuperscript{th} February 2025}

\pubyear{2025}

\begin{document}
\label{firstpage}
\pagerange{\pageref{firstpage}--\pageref{lastpage}}
\maketitle

\begin{abstract}
We analyse the 2021 outburst from the black hole X-ray binary \gx\ observed by \nicer\ around the hard to soft transition, when the system exhibits flip-flops between two distinct luminosity states: a bright state with a 5-6\,Hz quasi-periodic oscillation (QPO) and a dim state showing only strong broadband noise. Despite the marked differences in variability patterns between these states, the spectral energy distributions remain strikingly similar, with only minor changes in the black body component in the soft X-ray range. We find that the QPO frequency correlates with the X-ray count rates and hardness, suggesting a tight coupling between the QPO mechanism and the accretion disc's spectral properties. Additionally, we demonstrate that flip-flops can occur on very short timescales, with almost 50 state changes within $\sim1200$\,s, while both states can also remain stable over longer periods (at least $\sim1000$\,s). We explore various QPO models to explain these observations, including the possibility that the corona's accretion speed is near the sound speed, affecting the presence of QPOs. However, the exact mechanism driving the flip-flops and the QPOs remains unclear. Our findings emphasize the complexity of these phenomena and the necessity for further theoretical and observational studies to unravel the intricacies of QPO and flip-flop behaviours in X-ray binaries.
\end{abstract}



\section{Introduction}

Black hole X-ray binaries are binary systems composed of a stellar-mass black hole and a companion star. Such sources were discovered in the 1960s \citep[see, e.g.][]{1964Natur.201.1307B,1964ApJ...140..460M}, and rapidly showed rich phenomenology in both X-ray \citep[e.g.][]{1967Natur.216..773F} and radio \citep{1968Natur.218..855A, 1971ApJ...168L..21H}. Nowadays, we know dozens of such sources that undergo major outbursts throughout their lives, with X-ray luminosities increasing by many orders of magnitude \citep{2006ARA&A..44...49R, 2010MNRAS.403...61D, 2016ApJS..222...15T}. During an outburst, a source may present two major different accretion states: (1) a hard state and (2) a soft state. These states are originally defined by their X-ray spectral shape but also have characteristic fast variability properties. The hard-state is characterised by a power-law component that dominates the X-ray emission, from $1$ to more than $100$\,keV, with a typical index $\Gamma \sim 1.5-2$. This hard X-ray emission is emitted by a region whose geometry remains debated to this day, and we will simply refer to it using the agnostic term `corona'. In turn, the soft state exhibits a strong thermal component, usually well-fitted by a black body with a temperature around $1$\,keV, accompanied by a much weaker power-law component that usually represents a few percent of the flux with an index $\Gamma \sim 2.5$. During an outburst, the source usually goes through both states, along with major changes in the overall flux. However, the reasons for such behaviours are still debated \citep{2007A&ARv..15....1D,2014ARA&A..52..529Y}.

In the present work, we focus on the variability observed in the X-ray band. Such variability has been studied since the 1980s \citep[see, e.g.][]{1989ARA&A..27..517V}, particularly since the launch of the \textsc{Rossi X-ray Timing Explorer} in December 1995, and more recently the \textsc{Neutron Star Interior Composition Explorer Mission} (\nicer) in June 2017. Using these instruments, coupled with mathematical techniques such as (fast-)Fourier transforms, one can probe the variability of X-ray binaries in the X-ray band. Of particular interest is the power spectrum, i.e., the intensity of the Fourier transform as a function of frequency. 
During the outbursts from X-ray binaries, the power spectrum can exhibit narrow peaks called  quasi-periodic oscillations (hereafter QPOs) and fit using Lorentzians (see Appendix~\ref{app:psfitting} for notation conventions); providing a characteristic frequency $\nu$ and width $\Delta \nu$ (or $Q$-factor defined as $Q=\nu/2\Delta\nu$).
For a peak to be narrow enough to be classified as quasi-periodic, $Q\ge2$ \citep{Nowak2000}.
There are different classes of peaks that can be studied, most primitively divided by frequency at $\sim50$\,Hz. We focus here on the low-frequency ones. These can have different characteristics and properties \citep[e.g.][]{1997A&A...322..857B,2002ApJ...572..392B,1999ApJ...526L..33W,2001ApJS..132..377H,2003A&A...412..235N,2011MNRAS.418.2292M,2015MNRAS.452.3666S,2015MNRAS.451..475U,2015ApJ...808..144K,2017ApJ...845..143Z}, and are divided into three classes: type A, B and C \citep{2005ApJ...629..403C}.
Type A QPOs usually appear at the end of the hard-to-soft transition. They are quite rare and weak, with a few percent rms and a broad peak $(Q \lesssim 6)$, and have a typical frequency around $6$\,Hz. Type B QPOs are much more common and narrow $(Q \geq 6)$ than type A. They are usually observed around $\nu \approx 1-6$\,Hz, and are commonly detected before type A; i.e., during the transition from the hard state to the soft state. Finally, type C are the most commonly detected class of low-frequency QPOs. They cover a wide range in frequency, from mHz to $\gtrsim 10$\,Hz, are usually narrower than Type B $(Q \geq 10)$, and are observed throughout the hard state (but may also be present in the soft state). The main difference in the power spectrum between times including a type A/B and type C QPO is the presence of a strong broad-band noise (hereafter BBN) along with the type C QPO. This BBN is usually detected at frequencies below that of the QPO \citep[see, however][]{2024MNRAS.527.5638Z} and can account for a significant fraction of the variability.
There are several candidate models to explain the behaviour of low-frequency quasi-periodic oscillations around X-ray binaries, and we invite the reader to reviews such as \citet{2019NewAR..8501524I} or \citet{2022hxga.book...58D}.

In addition to these narrow components in the power spectrum, there are variability features that are not as commonly studied: Dips and Flip-flops.
Dips refer to a sharp decrease in the count rate \citep[e.g.][figure 4]{2004ApJ...610..378P}. Such dips are often understood as being due to absorption features, covering parts of the accretion flow \citep{1987A&A...178..137F, 1982ApJ...253L..61W}. Due to the presence of two states (absorbed/unabsorbed), they are often associated with flip-flop events. Flip-flops were first reported by \citet{1991ApJ...383..784M} for GX 339-4, and they remain an infrequent occurrence in black hole transients, observed only in a handful of sources; see, for example, XTE J1859+226 \citep{2004A&A...426..587C}, XTE J1817-330 \citep{2012A&A...541A...6S}, Swift J1658.2–4242 \citep{2020A&A...641A.101B}. They resemble top-hat-like variations in X-ray flux, where the source transitions between a low-flux (dim state) and a high-flux (bright state) level over timescales ranging from a few seconds to a few days. The major difference with dips is the stable character of both states. During a dipping event, the source lies in the bright state and progressively and briefly drops to a dimmer flux. However, during a flip-flop event, the source spends significant time in both edge states and very little moving between them. Moreover, while the flux can change by a significant fraction (up to $77\%$, \citealt{2020A&A...641A.101B}), the overall shape of the spectral energy distribution is not significantly impacted \citep[e.g.][]{2023MNRAS.521.3570Y}. In addition to the change in flux, the variability pattern of the source can also change between both states during flip-flops: states can sometimes each exhibit a different type of QPO \citep[e.g.][]{2004A&A...426..587C,2020A&A...641A.101B}, no QPOs at all (e.g. \citealt{2011ApJ...731L...2K,2022MNRAS.513.4308L}\footnote{This case has also been labelled alternating flux states (AFS) due to the absence of change in the variability pattern.}), or be consistent with the appearance/disappearance of a QPO (e.g. \citealt{2021MNRAS.505.3823Z,2023MNRAS.521.3570Y}). Adding to the complexity, different behaviours have been observed within the same source. For example, in GX339-4, \citet{1991ApJ...383..784M} observed flip-flops associated with the (dis)appearance of a type B QPO in its 1988 outburst, while \citet{2022MNRAS.513.4308L} report variability patterns during a 2021 outburst that resemble those of a soft state, i.e., no QPOs.

The literature on flip-flops and related phenomena in accretion disks around black holes highlights several key insights and hypotheses. For example, \citet{1997ApJ...489..272T} suggest that flip-flops may result from transitions between thermally unstable and viscously unstable disk states. In turn, \citet{2003A&A...412..235N}, while uncertain whether their observations are flip-flops or dips, emphasizes the need for theoretical models to account both for the sharp appearance and the rapid variability of QPOs. More recently, \citet{2013A&A...552A..32K} attribute transition dips to instabilities in the inner accretion flow, though the exact nature of these instabilities remains unidentified. More recently, \citet{2019ApJ...879...93X} associate rapid flux variations with transient QPOs, proposing that these are due to fundamental changes in the inner accretion flow driven by instabilities originating at larger radii. These instabilities then propagate inward, affecting the disk's global structure and altering X-ray spectral and timing properties. Together, these studies underscore the complexity of accretion disk dynamics and the critical role of instabilities in driving observed variability.

Moreover, the connection between flip-flops and Quasi-Periodic Oscillations (QPOs) in accretion disks around black holes is an area of active research. \citet{2022MNRAS.513.4308L} study flip-flop like patterns in the light curve of \gx\ and note that these diverge from typical observations as these transitions occur before the soft state and are associated with strong reflection features not commonly seen in flip-flops. Interestingly, this study does not observe any QPOs, suggesting that QPOs may not be inherently linked to flip-flops but could be a by-product. In a similar vein, \citet{2022ApJ...938..108L} observe flip-flops with type B/C transitions and propose that the disappearance of QPOs might be due to the disk aligning with the black hole's spin. In turn, \citet{2019ApJ...879...93X} provide another perspective, indicating that coronal properties remain invariant during flip-flops despite significant flux variations. This invariance suggests that the corona does not change during flip-flops, implying that QPOs should also remain unaffected. This supports the idea that QPOs are not directly linked to the core processes driving flip-flops. Lastly,  \citet{2012A&A...541A...6S} report that in the source XTE J1817-330, the transition from type-B to A QPO occurs on a short timescale and could be related to variable jet or outflow onset in the accretion disk. This finding highlights the potential role of jets or outflows in influencing QPO behaviour during flip-flop transitions.

Collectively, these studies suggest that while there is an association between flip-flops and QPOs, the exact nature of this relationship remains complex and may involve multiple factors, including the alignment of the disk with the black hole spin or the behaviour of the jets/winds. Given the current lack of understanding surrounding the flip-flops and QPOs phenomenon, our research aims to delve deeper into this area. We have chosen to focus on the 2021 outburst from \gx, one of the most extensively studied sources in the literature, for several reasons. Firstly, \citet{stiele23} reported unusual power-density spectra during this outburst, highlighting its peculiar variability nature. Secondly, \citet{2022MNRAS.513.4308L} observed the source a few days after  \citet{stiele23} and noted a total absence of QPOs during some flip-flops. We aim to investigate this specific period to improve our understanding of the underlying mechanisms responsible for both flip-flops and QPOs.

The present paper is organized as follows. In Section~\ref{sec:obs}, we describe the observations and standard data analysis methods used (some description of more bespoke methods is given in Section~\ref{app:psfitting}). We describe the observational phenomena in Section~\ref{sec:results} and how they may be interpreted by current theory in Section~\ref{sec:discussion}. Finally, we summarise our conclusions in Section~\ref{sec:conclusion}.

\section{Data analysis}
\label{sec:obs}

We study \nicer\ observations from around the state transitions of \gx\ during its 2021 outburst (the previous full outburst of \gx\ during the \nicer\ era occurred around Sun constraint, so was not well observed). Based on the analysis by \citet{stiele23}, SK23 hereafter, we select OBSIDs 4133010103-8 as covering the hard-to-soft transition and 4133010251,2,5,6,7; 4675010701 as covering the soft-to-hard (OBSIDs 4133010253,4 contain no good time). We reduce the data using the standard \nicer\ pipeline as of \textsc{heasoft} 6.32.1. We leave all screening parameters at their default values, which gives a high data yield with no apparent instrumental features.

Unless stated otherwise, we calculate periodograms from light curves with a resolution of 1\,ms, using segments of 8.192\,s ($2^{13}$ bins).
Long-term \nicer\ light curves are binned to the complete segments used to calculate each periodogram.
For clarity when plotting, we apply further visual binning as follows (though statistical calculations are made on the raw products).
Dynamic power spectra are smoothed over $3\times3$ pixels in adjacent frequencies and segments (in some cases, further binning may occur for readers due to the limits of display/printer resolution).
Power spectra are binned in frequency space, with each bin having the minimum width required to give a signal to noise ratio $\ge5$, subject to  $0.05\le\Delta\nu/\nu\le0.2$.

We also show long-term light curves from \maxi\ \citep{matsuoka09}, using the provided online light curves\footnote{http://maxi.riken.jp/star\_data/J1702-487/J1702-487.html}.
For clarity of plots, we impose additional filters on the rate uncertainty of at most 0.1\,photons\,cm$^{-2}$\,s$^{-1}$ and at most half the net rate.
Furthermore, when plotting hardness, we use the alpha (opacity) colour channel to emphasise higher quality observations, as $\alpha=\frac{h}{5\sigma_{\rm h}}$ (clipped to the interval $[0,1]$).

\begin{figure*}
    \centering
    \includegraphics[width=0.99\textwidth]{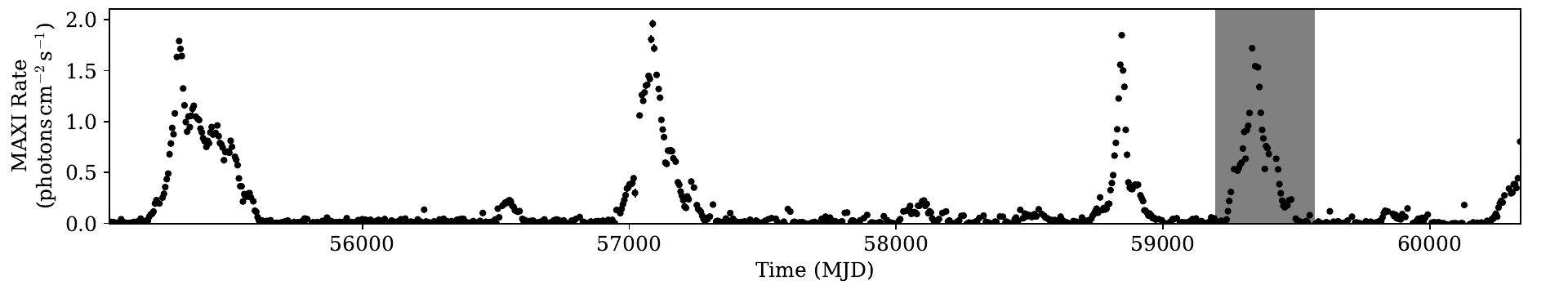}
    \includegraphics[width=\columnwidth]{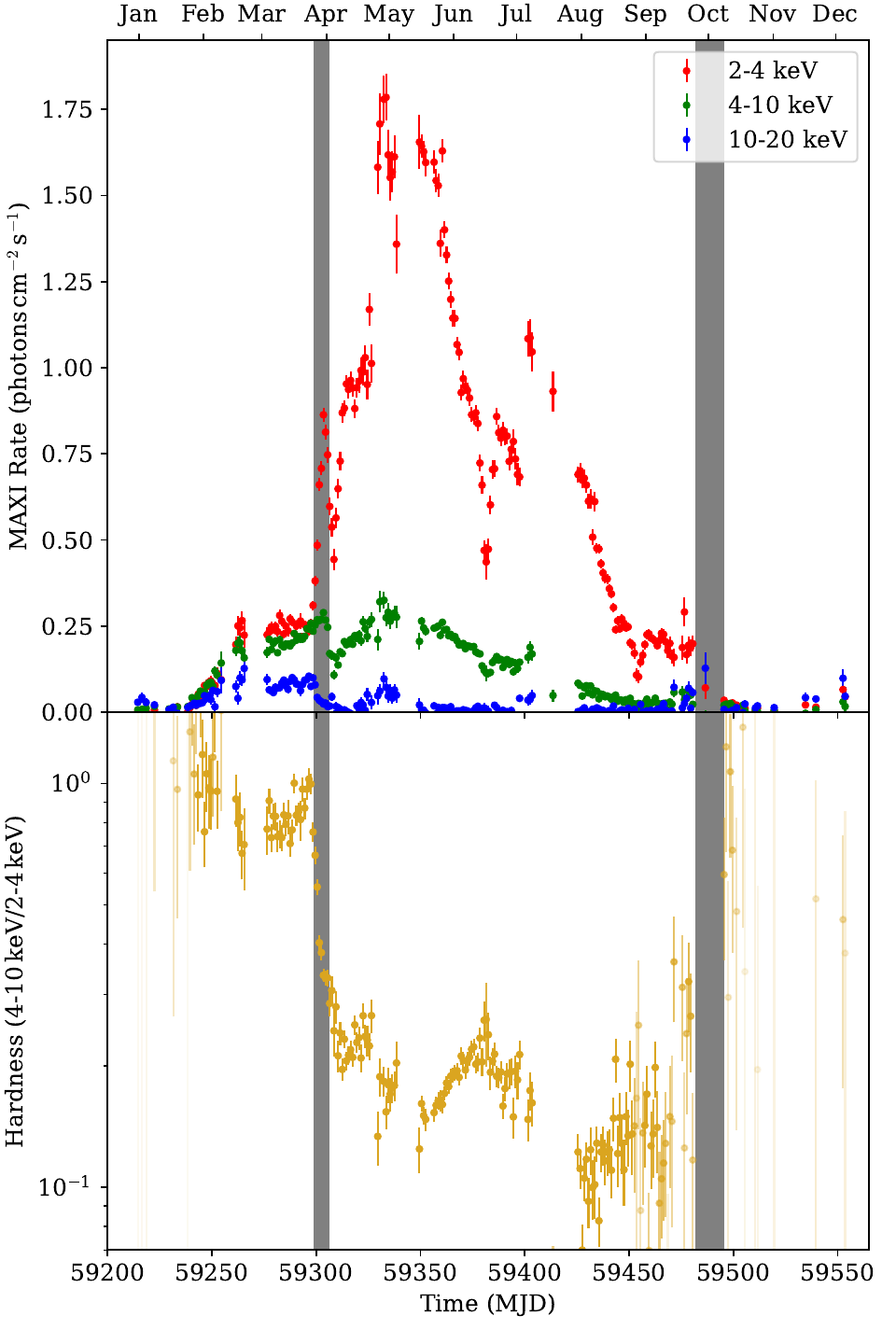}
    \includegraphics[width=\columnwidth]{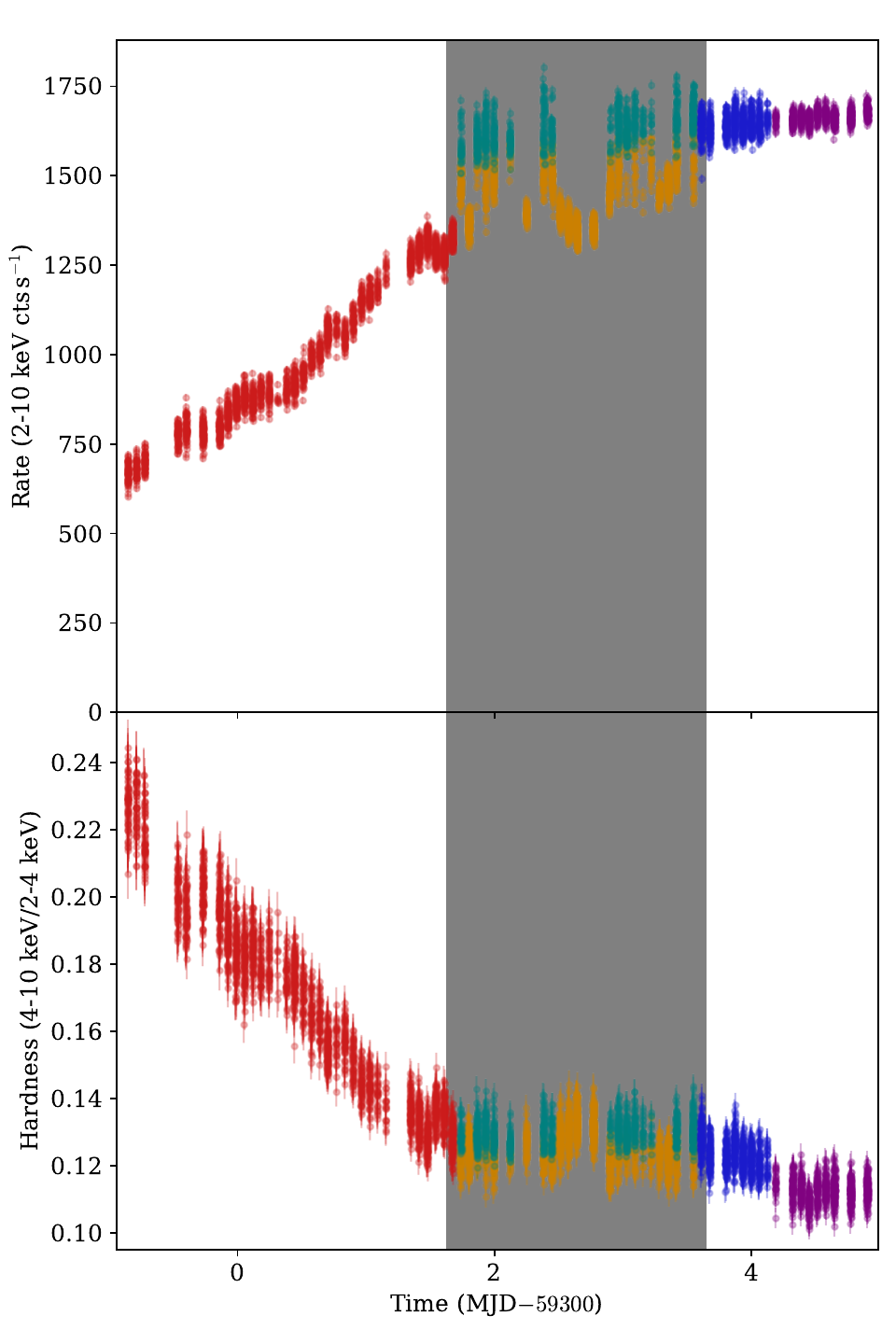}
    \includegraphics[width=0.99\textwidth]{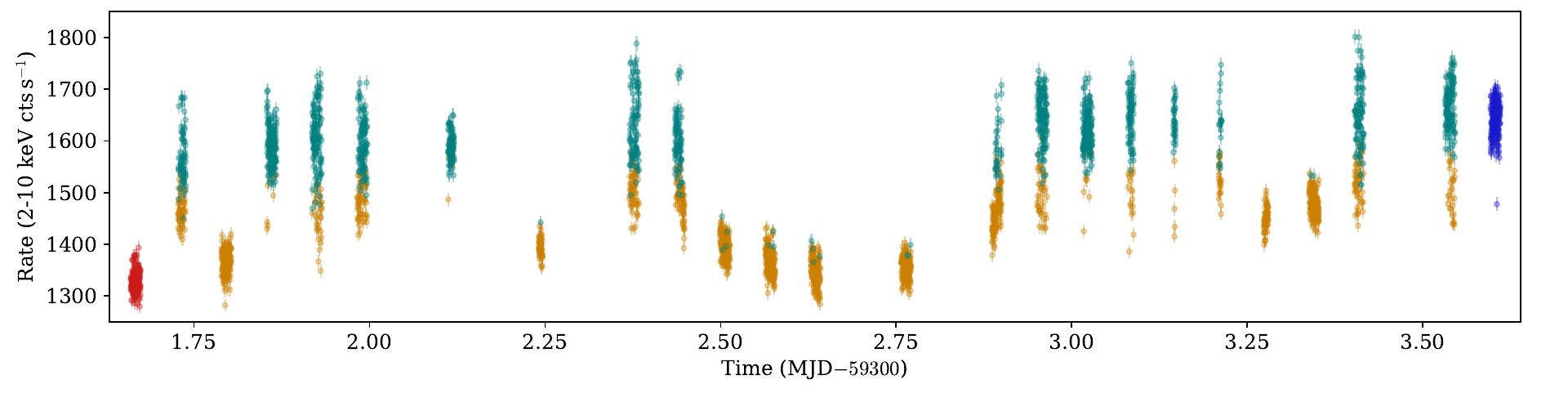}
    \caption{Light curves of \gx: {\it Top:} the long term 2-20\,keV light curve from \textit{MAXI}, showing several large and several smaller outbursts. The time interval in the {\it left} panels is shown shaded here. {\it Left:} The \textit{MAXI} light curves ({\it upper}, coloured by energy band) and hardness ratio ({\it lower}) of the outburst considered here. The times of the forward transition shown in the {\it right} panels and the reverse transition described in the text are shown shaded. {\it Right:} \nicer\ light curves and hardness ratios covering the state transitions. Colours mark the divisions described in the text: red before the hard to soft transition; cyan and yellow during; and blue and purple after. The time range shown in the {\it bottom} panel is shown shaded {\it Bottom:} The transition region of the light curve from \nicer.
    }    
    \label{fig:lcs}
\end{figure*}

\section{Results}
\label{sec:results}

\subsection{Long-term properties}

\begin{figure}
    \centering
    \includegraphics[width=\columnwidth]{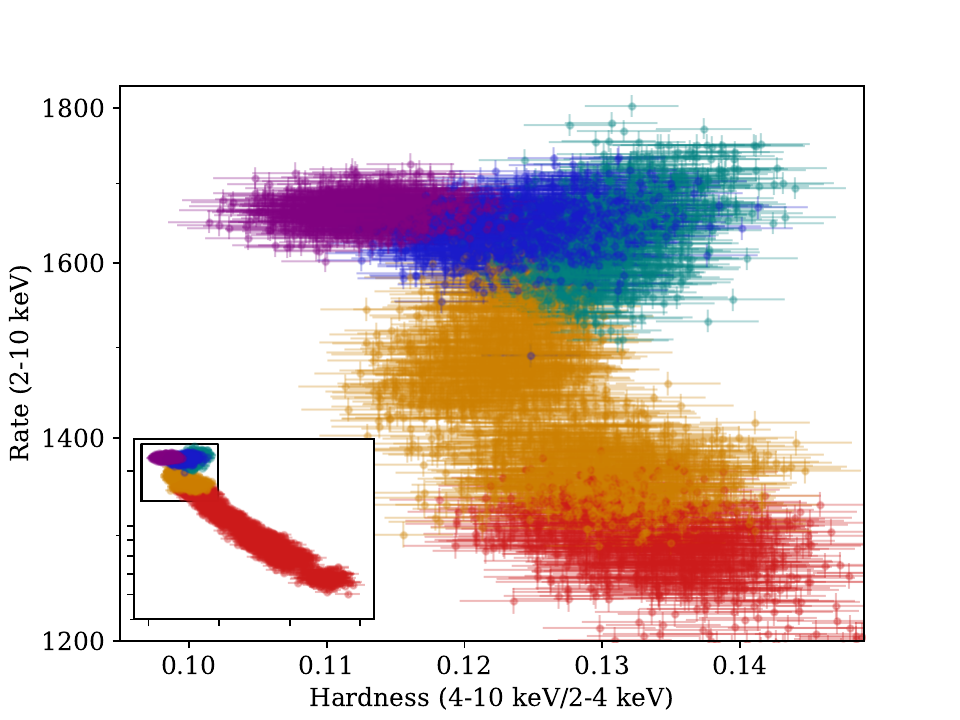}
    \caption{The hardness-intensity diagram of the \nicer\ data covering the forwards state transition. Colours mark the divisions described in the text. All four segments are well separated in hardness-intensity space, despite the middle two being well mixed in time.}
    \label{fig:hid}
\end{figure}

We begin (Figure~\ref{fig:lcs}) by showing the long-term light curves of \gx, over the $\sim15$\,years covered by \maxi. The outburst considered here is fairly typical of a full outburst of \gx, with four complete and one ongoing having occurred during the \maxi\ mission (there are also several smaller, hard-only or `failed' outbursts, where the source remains fainter and in the hard state, e.g. \citealt{2021MNRAS.507.5507A,2021NewAR..9301618M}).
Within the outburst we consider, we see that the transitions are typical of the canonical picture \citep[e.g.][]{2006ARA&A..44...49R,belloni16} of X-ray binary outbursts and similar to other full outbursts of \gx: the emission softens rapidly during the bright interval of the outburst; and it hardens again after fading.

Considering the forward transition in more detail, we see an interval of enhanced medium timescale ($10$-$10^4$\,s) variability (cyan/orange) where the observed count rate oscillates on moderately long ($>1000$\,s) timescales and  an associated kink in the track in hardness-intensity space (Figure~\ref{fig:hid}).
This interval has previously been associated with various rapid changes in timing properties \citep[e.g.][]{2022MNRAS.513.4308L}.
Foreshadowing our later findings, we will refer to this interval as the (state) transition.

Based on these gross light curve properties, we divide the emission into 5 sections: the hard state before the transition (red); a bright (cyan) and a dim (orange) section during the transition; and two sections (blue/purple) after the transition (this final division foreshadows results later in the paper).

\subsection{Spectral energy distribution} \label{sec:SED}

\begin{figure}
    \centering
    \includegraphics[width=\columnwidth]{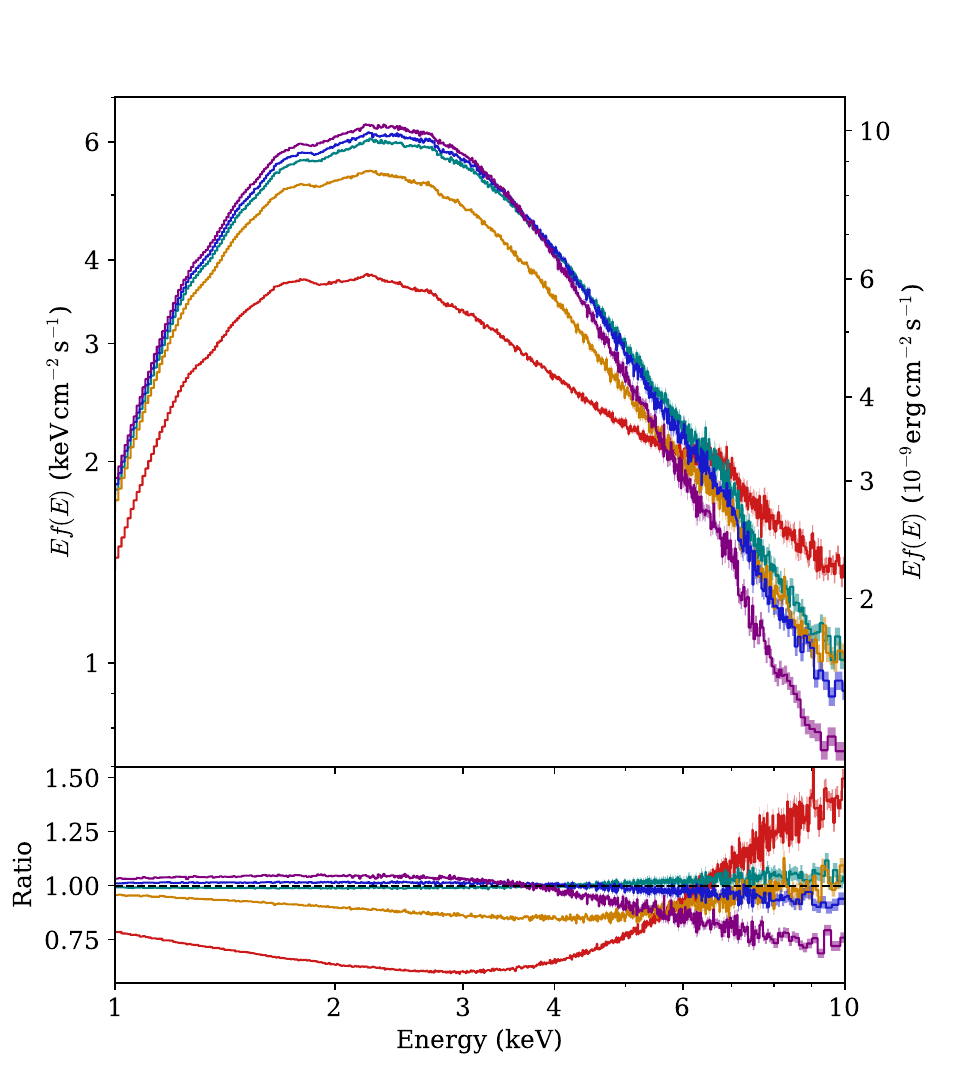}
    \caption{Top: Mean spectral energy distributions of each state (not corrected for foreground absorption). Bottom: Ratio of given state compared to the mean of the bright and soft-intermediate states (cyan and blue, see text). In both figures, the states are shown in their usual colours: red for hard-state, orange for dim state, cyan for bright state, blue for soft-intermediate, and violet for soft-state.}
    \label{fig:sed}
\end{figure}

In Figure~\ref{fig:sed}, we show the mean spectral energy distribution (SED) of each section, unfolded as the ratio to a $\Gamma=2$ power law.
While the hard-state is quite different, in part because it is the average of a much broader sets of spectra (see Figure~\ref{fig:hid}), the other four spectra are extremely similar both in flux and general shapes. All the considered states seem to be disk-dominated, with the emission peaking at $2-3$\,keV, confirming the best-fit temperatures found in the literature ($0.7-0.8$\,keV, \citealt{2023MNRAS.521.3570Y}). We show on the bottom panel the ratio of each spectra compared to what we think would usually be considered the soft-intermediate state, the average between the cyan and blue portions. This figure confirms a few expectations.
First, the hard-state has a stronger power-law component (i.e., more coronal emission) than other states, as suggested by its higher hardness. Its most significant difference to the other four states lies above $6$\,keV, illustrating that this is where the emission begins to be dominated by the corona\footnote{See also Insight-HXMT data from the same period \citep{2023ApJ...950....5L}.}.
Second, the evolution from the bright (cyan), through the soft-intermediate (blue), and until the soft (violet) state is quite smooth and due to a simple softening of the source: the disk (below $6$\,keV) becomes brighter and the power-law component (above $6$\,keV) fainter.
Third and more importantly, this coronal emission changes little (at least for energies up to 10\,keV) between the dim (orange) and bright (cyan) states. In fact, even the emission below $6$\,keV only differs by a mere $10 \%$ at most, see bottom panel.

The lack of significant change in the spectral energy distribution, in particular in the power-law component, is a strong evidence that the properties of the corona have not substantially changed. Note that we here only rely on \nicer\ data, but the lack of change in the coronal properties can be seen directly by using a harder X-ray range \citep[e.g.][figure 2]{2023MNRAS.521.3570Y}\footnote{Note that these authors' definition for the flip-flop location in time is different from ours.}. 

\subsection{Fast variability}
\label{sec:DPS}
\label{sec:psd}
\label{sec:meanPDS}

\begin{figure}
    \centering
    \includegraphics[width=\columnwidth]{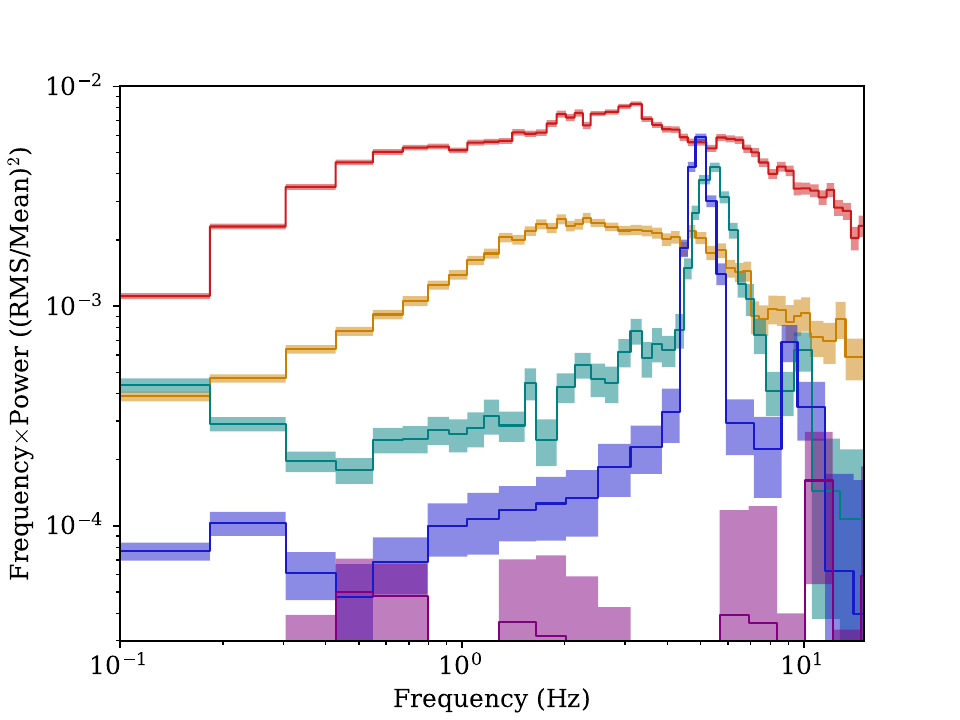}
    \caption{Mean power spectrum for each state. The QPO is not visible in the hard-state (red) due to its large frequency change during the measured interval.} 
    \label{fig:psds}
\end{figure}

\begin{figure*}
    \centering
    \includegraphics[width=\textwidth]{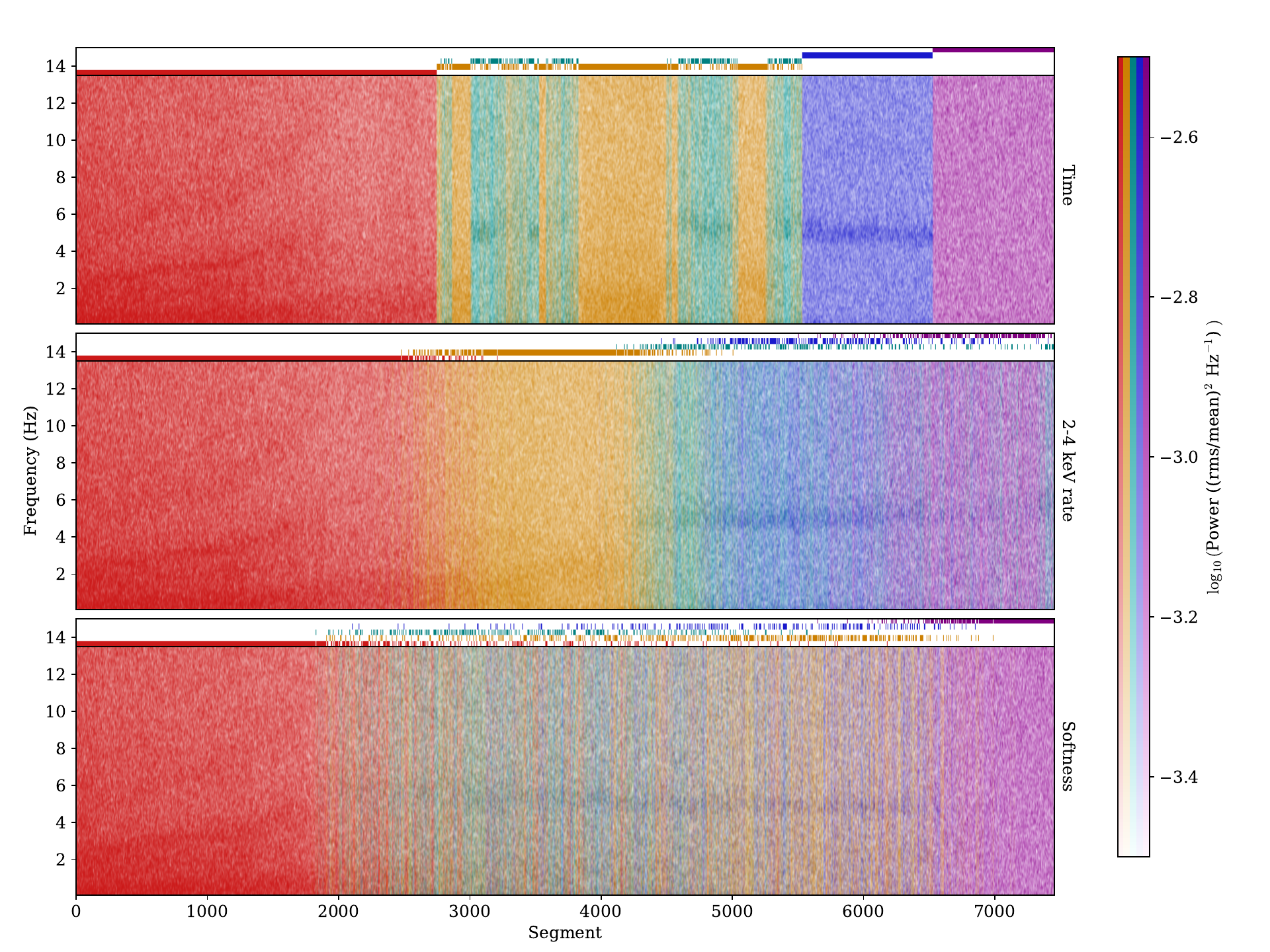}
    \caption{Dynamical power-spectrum of the observations chosen in this paper. Top: sorted by time, with time gaps removed. Middle: sorted by observed soft ($2-4$ keV) count rate. Bottom: sorted by softness, i.e. ratio of soft ($2-4$ keV) to hard ($4-10$ keV) count rates. On all panels, segments are $8$ seconds long and the different states are shown in their associated colors (Figure~\ref{fig:hid}).}
    \label{fig:dynamic_psds}
\end{figure*}

Based on the spectral properties (and, to a lesser extent, long-term variability) of the source, we have divided the emission around the state transition into five sections (although the final division is motivated by a change in the fast variability, so will be justified later in this section).
Next, we consider how these divisions compare with the properties of the fast variability.

In Figures~\ref{fig:psds}~and~\ref{fig:dynamic_psds}, we show the mean (per state) and dynamic power spectra of the complete data set under consideration. We show power spectra for  2-10\,keV throughout, since the results for different bands within the \nicer\ range are similar (although the overall variability power in higher energy bands, with more coronal emission, is higher, as is typical).

The mean power spectra in Figure~\ref{fig:psds} show the large change in variability amplitude and frequency distribution across the transition.
Variability is greatest during the pre-transition interval (red), consisting of a broad peak with some wiggles due to averaging of different power spectra (as we show in the next section).
During the flip-flops, the dimmer segments (orange) have a similar power spectrum to the pre-transition spectrum, but with lower amplitude.
The brighter segments (cyan) are very different, with much lower broadband variability and a QPO at $\sim6$\,Hz.
After the transition interval, the power spectrum is similar to that of the bright intervals but the broadband variability is slightly lower still and the QPO appears sharper (blue). Again, as we will show later, this is due to averaging slightly different power spectra during the transition.
The final section (purple) has minimal variability, with the QPO also having disappeared.

\subsubsection{Time sorting}

In the top panel of Figure~\ref{fig:dynamic_psds}, we see that the hard interval has QPOs that evolve in frequency and strength; the transition interval has grossly different power spectra in the bright and dim segments; and the post-transition interval initially shows a steady QPO, which then disappears.
At the start of these observations, the source lies in the hard-state (red), where there is a clear QPO whose frequency increases from $2$ to $4-5$ Hz (first 2000 segments) before it fades. A broad-band noise (BBN) component is present throughout this entire hard state. Then, around segment 2700, the source enters the flip-flop states, alternating between the bright (cyan) and dim (orange) modes. The bright state includes a clear QPO with a rather constant frequency of $5-6$ Hz, and any broad-band noise is low. Contrarily, in the dim state, the QPO has now disappeared and BBN is clearly visible. Around segment 5500, the flip-flops seem to have ended and we enter what seems to be a soft-intermediate phase (blue), where no BBN is detected and a QPO is observed, with a frequency around $6$ Hz.
Finally, the QPO disappears as the source reaches the soft state interval (violet), where the variability is at the lowest (no QPO, no BBN).

The overall behaviour shown in this panel is quite erratic, especially during the flip-flops. They can have widely different duration, with sections where a dim or bright interval (cyan/orange) is inhabited for only a few consecutive segments (around segment 3500), while other intervals last for hundreds of segments (see between segments 4000 and 5000), i.e., thousands of seconds. Note that thousands of seconds are millions of dynamical timescales in the inner regions; in other words, each configuration (dim or bright) has to be stable.
After the highly variable part of the transition, the power spectrum remains similar to the bright intervals during the transition, consisting of a powerful QPO and much lower level of noise.
Finally, the QPO vanishes sharply after some significant time in an apparently steady state.

Given the similarity of the pre/dim and bright/late pairs, we also sort the segments forming the dynamic power spectrum in a way that naturally places these intervals adjacent to each other.
While there is no physical grounds to perform such re-organization (yet), we believe that we can realistically perform such study because the segments considered here are much longer than the expected dynamical timescales (see section~\ref{sec:QPOmodels}).

\subsubsection{Rate sorting}

Following the idea that QPO properties are tightly linked to the count rate of the source \citep[e.g.][]{2003A&A...412..235N}, we sort segments by soft (2-4\,keV) rate (Figure~\ref{fig:dynamic_psds}, middle panel). In this panel, the state with the lowest rate is positioned on the left and the highest on the right. As before, the evolution starts with the hard-state (red), and ends with the soft-state (violet), however, the middle part has been restructured by soft count rate.
The flip-flopping states, dim in orange and bright in cyan, are not alternating any more. Instead, there is now a transition from the dim state to the bright state, consistent with their definitions. The dim state seems to seamlessly connect with the hard state (red-to-orange), while the bright state connects with the soft-intermediate state (cyan-to-blue)\footnote{Note that the soft state (violet) is smeared within the soft-intermediate (blue) due to their similar count rates, and a more detailed physical separation would realistically distinguish them.}. Moreover, the QPO frequency seems to increase with the 2-4\,keV count rate throughout the whole observation; from 2 to 5\,Hz in the first portion (red/orange), and from 4.5 to 5.5 in the second portion (cyan/blue). More importantly, this figure illustrates a major characteristic of the flip-flops: when ordered by count rate, the dynamical power-spectrum draws a smooth picture of the variability. We make two contradictory observations from this image. First, the QPO and BBN seem to be unrelated. This is illustrated by the fact that all four possible variability states are observed: QPO and BBN (before segment 1700), BBN but no QPO (1700 to 4200), QPO but no BBN (4200 to 6500), and neither QPO nor BBN (after 6500). Second, despite the possibility that all four states are possible, there is a visual suggestion that the QPO appearing in the bright state (around 4200) emerges\footnote{See also the grey-scale version of this figure, see Figure~\ref{fig:dynamic_psds_grey}.} from the BBN. This is, of course, only an observation based on a re-organization of the segments, and is not a direct demonstration. Such a behaviour has already been indicated by previous studies, see for example \citet{2020ApJ...891L..29H} during the 2018 outburst from MAXI J1820+070. These authors show the appearance of a type B QPO (around $14000$ in their Figure 2) that seems to connect to the BBN which is contemporary to the disappearing type C QPO. However, the data is shown in compacted time, meaning that there are important time gaps in the data due to the observing pattern of \nicer. Equivalently, our data show this feature only after a spectrally based reorganisation of the dynamical power spectrum.

\subsubsection{Hardness sorting}

Given previous relations of QPO frequency with spectral hardness (4-10/2-4\,keV), we also sort segments by hardness (Figure~\ref{fig:dynamic_psds}, lower panel, sorted as softness to be more similar to the previous panels).
This panel shows these same features, though less clearly as the bright and dim sides of the transition region overlap significantly in hardness but not in rate (for these particular energy bands; compare the vertical and horizontal projections of the HID, Figure~\ref{fig:hid}).
This time, neither the QPO nor the BBN disappear before reaching the soft-state around segment 6500. Instead, the BBN seems to be present (yet, weak) throughout, and the QPO frequency increases at the beginning (until segments 1700-2000) before it slowly decreases until it disappears (around 6500); see section~\ref{sec:qpofreqevol} for a discussion on the frequency evolution. At first, this observation seems to be an artefact of the sorting used because the QPO (resp. BBN) is totally invisible in the dim state (resp. bright state), see bottom panels in Figure~\ref{fig:dpsd_separate}. Nonetheless, it is quite intriguing that the complex variability behaviour observed in both the time- (top) and rate-sorted (middle) is now smoother in both QPO and BBN evolution.

Therefore, we suggest that the two states of the transition interval are better regarded as extensions of the pre- and post-transition states than distinct states per se; instead, their distinction is only due to their overlap in time.


\subsection{Single observation with flip-flops} \label{sec:dynamic1orbit}

The behaviour shown in the previous sections is obtained by stacking data from several days.
However, the source can move between the bright and dim states in the light curve much more frequently, many times within a single \nicer\ snapshot.
To test whether the different variability behaviours also change on these timescales, we analyse a single \nicer\ snapshot displaying many bright/dim transitions;
we selected the snapshot around MJD 59302.37 (30 from SK23) as having the strongest rapid transitions.

\begin{figure}
    \centering
    \includegraphics[width=\columnwidth]{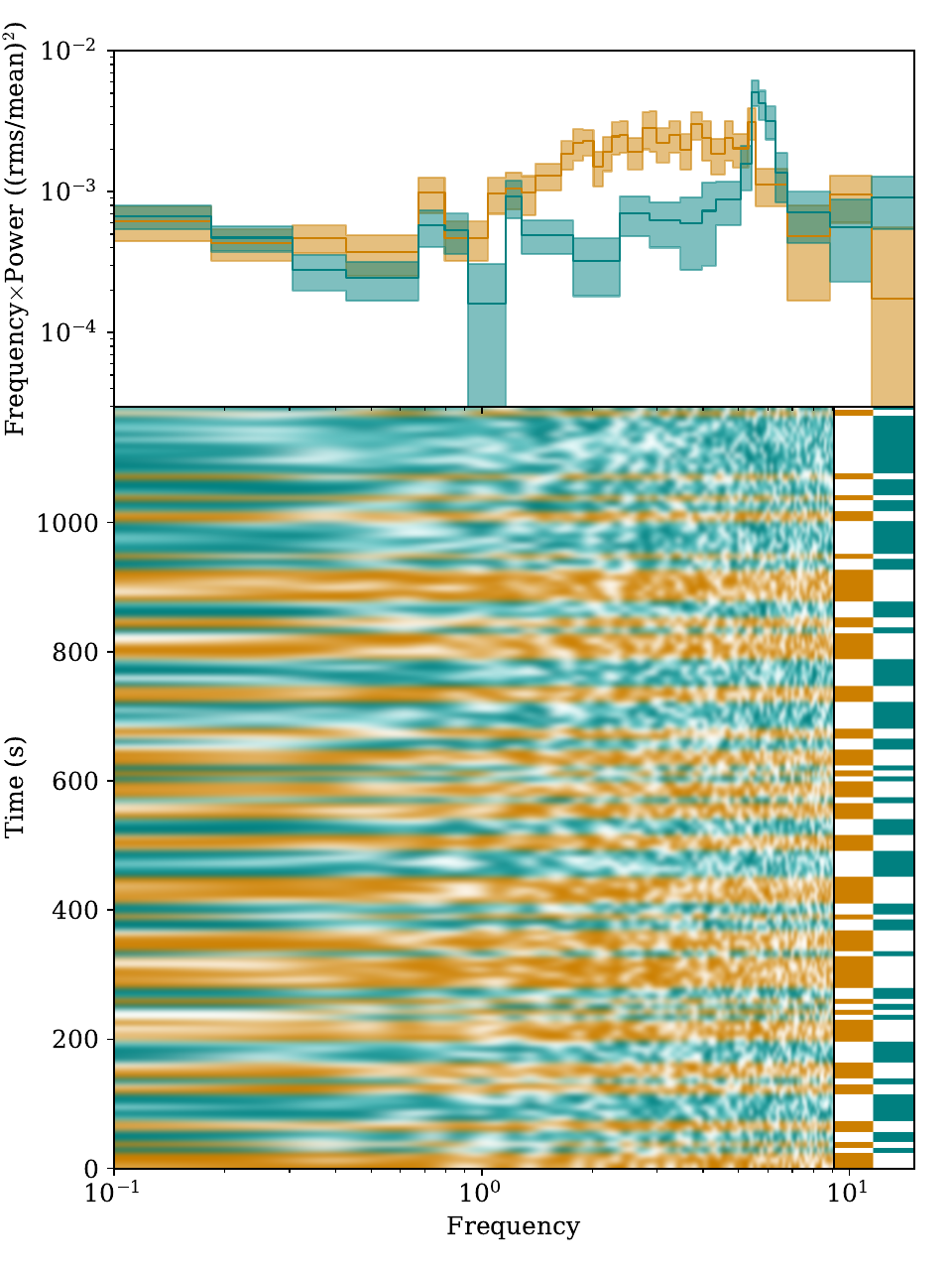}
    \caption{Mean (top, split by state) and dynamical (bottom) power spectrum of a single \nicer\ snapshot (around \nicer\ second 228474000). The colour scale of the lower panel is the same as Figure~\ref{fig:dynamic_psds}. The QPO and broadband dominated variability patterns can interchange on timescales down to a few tens of seconds.
    }
    \label{fig:dynamic1orbit}
\end{figure}

The \dps of the selected \nicer\ snapshot is shown on the bottom panel of Figure~\ref{fig:dynamic1orbit}. During most single snapshots throughout the flip-flops the source alternates between the two states, but in this particular snapshot it alternates between the states in a quite erratic way. Indeed, the full observation lasts about $1200$\,s, but there are almost $50$ state changes (i.e., one every $20-30$ seconds). Many of these intervals only last for a few seconds, which makes observing their variability an arduous task. Let us have a closer look at the longest intervals of each state. The longest dim interval lasts about $40$ seconds, and is located at around $900$\,s on the Y-axis. During this interval, one can see that the low-frequencies (below $3-4$\,Hz) dominate the power-spectrum, i.e., there is a strong BBN. In turn, the longest bright state lasts about $90$ seconds (located at $1100$\,s). In this case, one can perceive the presence of strong variability around $5-6$\,Hz, where the QPO is expected, as well as a much weaker BBN. A similar behaviour for either the dim or the bright state can be seen for other relatively long portions of this \dps .

To confirm this behaviour, we show the average power spectrum of each segregated portion, bright and dim, on the top panel of Figure~\ref{fig:dynamic1orbit}. Interestingly, despite the particularly erratic behaviour that led to the selection of this given observation, the two power-spectra are very discernible: the dim is clearly dominated by the BBN, while the bright undoubtedly exhibits a strong $5-6$\,Hz feature (i.e., a QPO) with a relatively weak BBN. This figure is a clear confirmation that the transition from dim to bright states can happen on a very short timescale. This behaviour during a single \nicer\ snapshot has been described before, see for example the top-left panel of Figure 4 in \citet{2022ApJ...938..108L}, studying the source MAXI J1348–630 during its 2019 outburst. Note that \citet{2022ApJ...938..108L} do not observe any BBN component during the dim state, which may be due to the particularly high energies chosen for the \dps ($35-100$\,keV, e.g. \citealt{1997A&A...322..857B, 2016ApJ...823...67S}).

There is lower frequency variability in the bright state around  $0.1$\,Hz, greater than that of the dim state. In fact, some states with a strong QPO show a stronger very low-frequency ($0.01-0.1$\,Hz) noise than others with no QPO, see for examples observations 19 and 23 from SK23, Figure B1. However, the state changes shown above happen every $20-30$ seconds, i.e., with a typical frequency around $0.03-0.05$\,Hz; the low frequency variability power that results from the rate changes between states will leak into the low end of the measured band. We address this question in the following section.

\subsection{Single observations without flip-flops} \label{sec:psd_adjacent}

\begin{figure}
    \centering
    \includegraphics[width=\columnwidth]{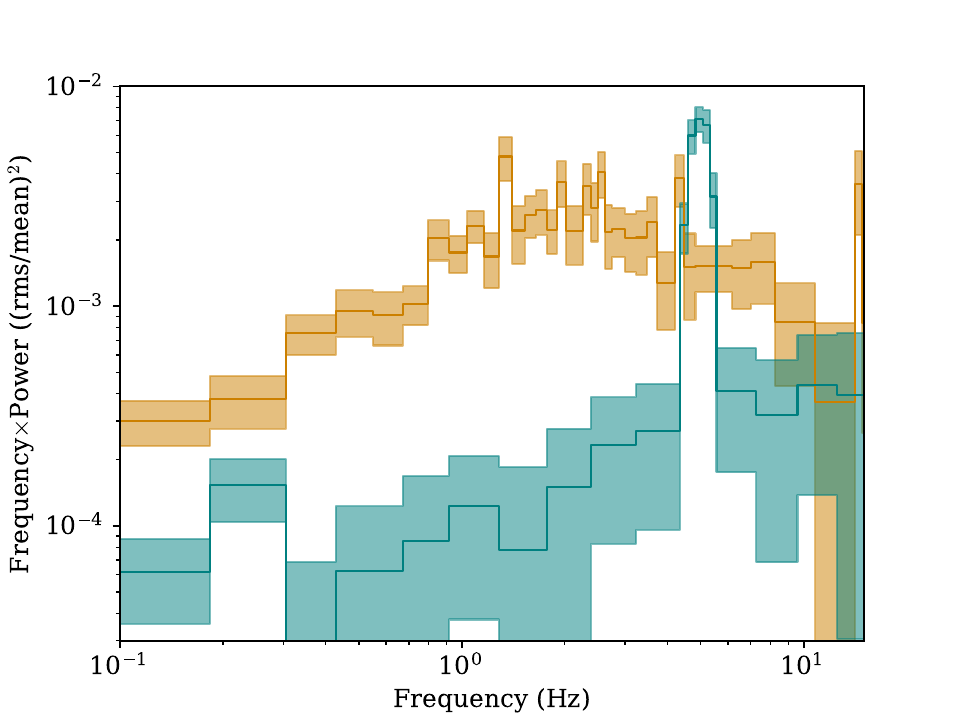}
    \caption{Power spectra of two adjacent snapshots (MJDs 59302.12, cyan, and 59302.25, yellow), each in a single state. Each power spectrum is similar to the mean for its respective state but the QPO is narrower.}
    \label{fig:psd_adjacent}
\end{figure}

We then select two single \nicer\ snapshots that follow one another in time; one with a pure bright state (MJD 59302.1, $\# 28$ from SK23), and one with a pure dim state (MJD 59302.25 , $\# 29$ from SK23). We show their power-spectra in Figure~\ref{fig:psd_adjacent}. As expected, there are no QPO features in the dim state, but the bright state displays a clear QPO with $\nu = 4.97 \pm 0.04$\,Hz and a width $\Delta \nu = 0.29 \pm 0.04$\,Hz, i.e. a narrow QPO with $Q \simeq 17$. Note that this QPO is much narrower than the one estimated with stacked data (section \ref{sec:psd}).

\begin{table}
    \centering
    \begin{tabular}{c|c}
         & Indices \\ \hline
        Pure bright state & 28 \\ \hline
        Pure dim state & 24, 29, 32 $\rightarrow$ 35, 42, 43 \\ \hline
        Dim+Bright & 23, 25, 26, 27, 30, 31, 36 $\rightarrow$ 41, 44, 45\\
    \end{tabular}
    \caption{State of observations shown in SK23 according to our classification. Note that all observations prior to \# 24 are in the hard state (red portion), i.e. are technically pure dim states, and all observations past \# 46 are in the soft-intermediate state (blue) and are technically pure bright states.}
    \label{tab:SK23}
\end{table}

Unlike the rapidly switching interval (Figure~\ref{fig:dynamic1orbit}), the low frequency ($\sim0.1$\,Hz) power is much lower in the bright state than the dim. This confirms our indication in the previous section that the low frequency power observed during the bright state was due to the state switching rather than being intrinsic to the bright state.
It is also noteworthy that this behaviour will have a stronger impact on the bright state because this particular state has more of its observed segments while the source is rapidly switching states. Indeed, out of the 23 different snapshots that displayed flip-flops, there are 8 pure dim state snapshots and only 1 pure bright (see bottom panel of Figure \ref{fig:dynamic_psds} and table \ref{tab:SK23}). This may be a statistical or a selection bias, but it is very puzzling that long steady intervals are more commonly in the dim state.
Nonetheless, the rapid transitions between the dim and the bright states are probably the explanation for the puzzling behaviour observed during many snapshots in the study presented by SK23. See for example snapshots $\# 23$, where one actually observes a blend of: (1) multiple phases of dim states, (2) multiple phases of bright states, and (3) the timing features resulting from the dim-bright transitions.

\subsection{Evolution of the QPO properties}
\label{sec:qpofreqevol}

Comparing the widths of the QPOs during the bright interval in the single snapshot in Section~\ref{sec:psd_adjacent} and Section~\ref{sec:psd}, we see that the QPO is locally narrower than the average over the full interval, implying that the full interval includes some change in QPO frequency (and/or intrinsic width).
Therefore, we now analyse in more detail how the QPO frequency changes.
Figure~\ref{fig:dynamic_psds} already showed that the QPO changes more smoothly with hardness or rate than with time, suggesting some physical link between the frequency and the (energy) spectrum.

During the bright intervals of the transition, where the QPO is present, sorting light curve segments by hardness or counts rate appears to show the QPO frequency increasing with the sorting parameter.
Therefore, we fit the periodograms of the individual light curve segments with a power spectral model that depends on these parameters. For comparison, we also fit the same model with segments sorted by time.
The details of the fitting method for individual periodograms sorted by some parameter are given in Appendix~\ref{sec:psdfitting}.
We use a PSD model consisting of a zero-centred Lorentzian for the broad band noise and a narrow Lorentzian for the QPO.
We then use a Bayesian prescription to derive an estimate of the posterior probability.
We use a flat prior on each parameter with a large initial range.
To estimate the posterior density, we use a MCMC chain \citep{2010CAMCS...5...65G,2013PASP..125..306F} with 10 walkers per parameter and 500 steps, burning the first 3000.
We find that including up to the fifth derivatives (in the formalism of Appendix~\ref{app:psfitting}) is required for the highest derivative to be consistent with 0.
The fitted values are given in Table~\ref{tab:qpo_pars5}.
All parameters of interest -- those describing the QPO shape -- are well constrained. The exception to this is the width of the noise component; this simply shows that the noise shape is not critical over the fitted range; the QPO parameters also do not correlate with the value of the noise width, so the lack of constraint is unimportant for these results.
Since each parameter is tightly constrained (of those that are of interest), in that the posterior has most of its support over a small range of parameter values, the exact prior shape over this range will have little impact.
Finally, we check that the distribution of observed values, relative to the best fitting model, is similar to that proposed in the fitting likelihood (Figure~\ref{fig:distTest}): the observed probability density is within 5\% for the vast majority of the distribution.

We show the inferred evolution of the QPO parameters with each sorting parameter in Figure~\ref{fig:psdFitPars}.
As expected, the QPO central frequency increases with hardness or rate in either band.
Additionally, the width increases with hardness but any evolution is less clear with rate.
Correspondingly, the quality factor is highest while the source is soft.
The change in QPO power is small, with slight increases with hardness of hard band rate.

While a model with no change in QPO frequency is a significantly worse fit to the hardness-resolved (or rate-resolved) data, we also fit such a model to the average PSD to demonstrate the effect on QPO parameters when not taking the hardness-frequency correlation into account.
In this case, we use the average of the periodograms used previously, with Gaussian errors, as is standard after averaging many periodograms. We use the same constant plus Lorentzian model, although the central frequency is now fixed rather than depending on hardness. Again, we make a Bayesian estimate of the posterior probability for each parameter with a flat prior on each; each parameter is again well constrained.
The resulting values are shown as an overlay on Figure~\ref{fig:psdFitPars}.
The fitted width of the QPO is notably larger, indicating that fits to average PSDs underestimate the quality factor of Type-B QPOs.
The QPO frequency is of course similar to the central value from the variable frequency fit simply due to the averaging across true QPO frequencies above and below this value.

In principle, we could perform a similar analysis on the changes of the Type~C QPO frequency. However, this QPO is not clearly detected around the transitions which are the focus of this work. Additionally, the larger changes and more complex broadband noise profile make the Taylor expansion method less appropriate.

\subsection{QPO-free intervals}

\begin{figure}
    \centering
    \includegraphics[width=\columnwidth]{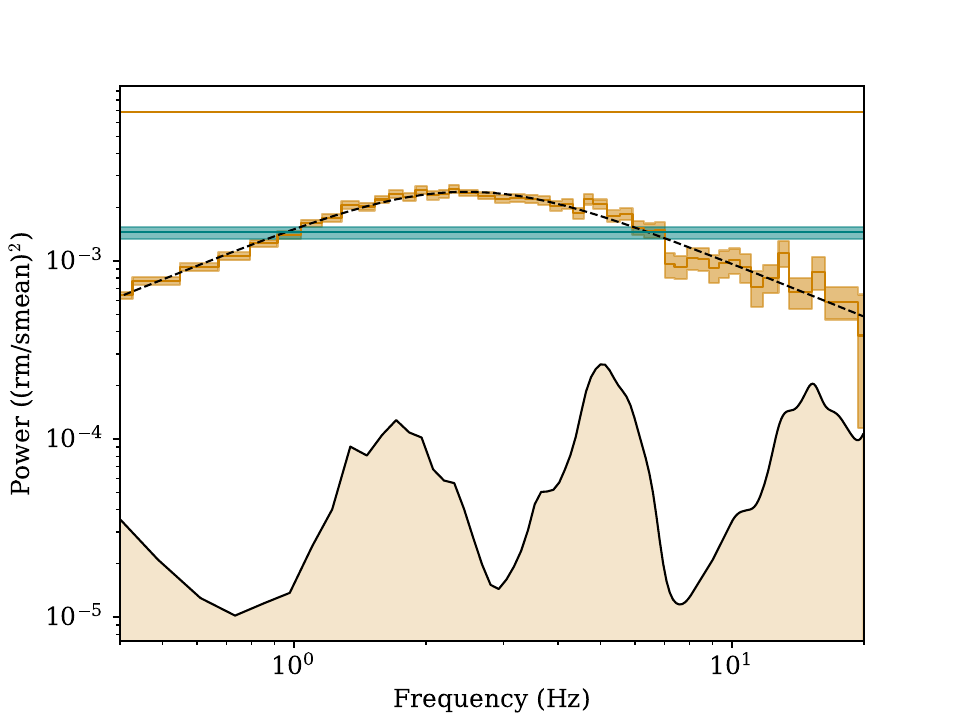}
    \caption{Upper limit (black solid) on power of a QPO in the mean power spectrum of the dim state, which is always significantly below the QPO power measured during the bright state (teal). For reference, we also show the integrated broadband power during the dim state (gold horizontal line); the measured PSD (gold, with shaded errors) and the QPO-free model (black, dashed).}
    \label{fig:upperlimit}
\end{figure}

Having shown that in some intervals the variability does not contain obvious QPOs, we now calculate upper limits for these stages.
There are many possible upper limits; we choose to show the limit for the mean PSD of the dim intervals as a function of QPO frequency.
The QPO width also affects the strength of a limit; for comparison with the bright intervals, we take $Q=6$, as the round number closest to the fitted value at the median value of the bright state (for any of the rate/hardness sorted fits). 
We then calculate an  approximate 90\% upper limit as a function of QPO frequency as the minimum QPO power which gives a fit worse than having no QPO by $\Delta\chi^2=2.706$. 
Figure~\ref{fig:upperlimit} shows that this limit is much lower than the QPO strength during the bright state, by at least a factor of $\sim3-30$ in power within the frequency range studied.

\subsection{QPOs during the reverse transition}

\begin{figure}
    \centering
    \includegraphics[width=\columnwidth]{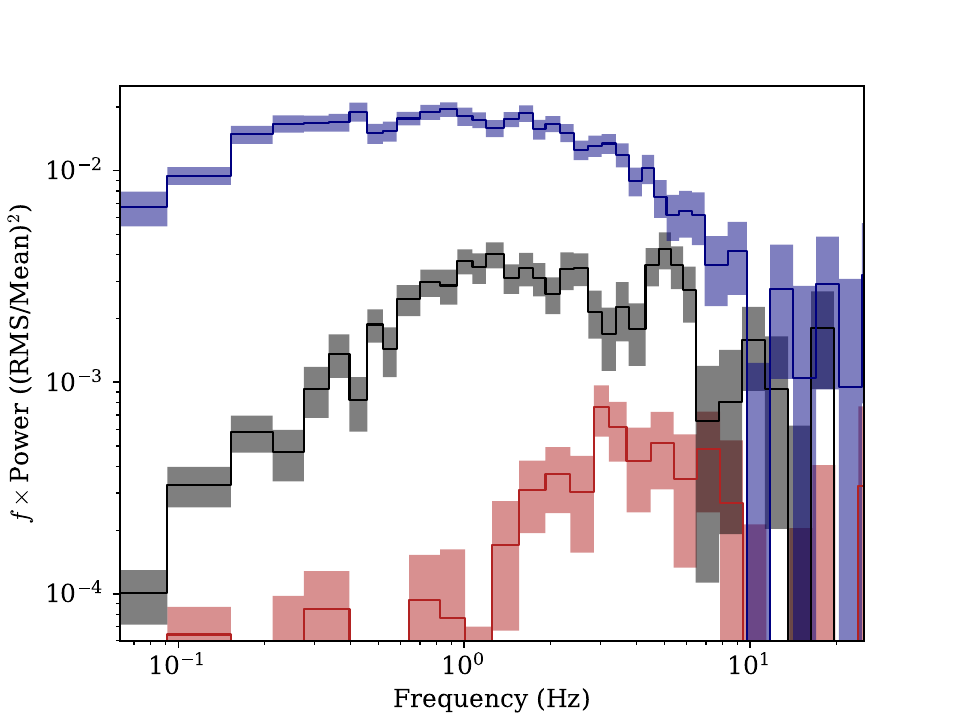}
    \caption{
    PSDs around the reverse transition. {\it Red:} before, OBSIDs 4133010251,2; {\it Black:} during, OBSIDs 4133010255,6; {\it Blue:} after, 4675010701,4133010257.
    During the reverse transition, there is a QPO similar to that seen in the forward transition. After the transition, there is broadband noise typical of the hard state, but no clear QPO.
    }
    \label{fig:reverse_psd}
\end{figure}

We also inspect the reverse (soft to hard) transition to analyse whether the QPO features described so far depend on the direction of transition. However, this transition was not well covered, as shown by the right panel of Figure~\ref{fig:lcs}.
Additionally, identifying the equivalent states to those described in the previous sections is not trivial:
because of the lower accretion rate and disc temperatures around this transition, observed count rates and hardnesses correspond to different physical conditions. Moreover, due to the lower count rate, (dynamical) power spectra are much noisier; therefore, we inspect the mean per (\nicer) snapshot power spectra.
These show a significant increase in variability -- the primary first-order state indicator -- across the likely transition time. We show in Figure~\ref{fig:reverse_psd} the power spectrum before (red; 4133010251,2), during (black, 4133010255,6), and after (blue, 4675010701,4133010257) the soft-to-hard transition studied here (numbers in brackets are the \nicer\ OBSIDs).

The variability before the transition is small (around $1\%$ rms), although there may be a broad peak around 2-3\,Hz, where the QPO will later appear. Across the transition, the variability increases and the average power spectrum now shows a clear peak around 5-6\,Hz, where the QPO was present during the hard-to-soft transition. After the transition, the power spectrum is similar to the long-term average seen during the initial hard state, showing strong broad band noise. Note that the data are not good enough to say whether a QPO is present but smeared due to a changing frequency (as it was during long-term averages of the hard state), or is absent from the data. 

We also attempt to find a split of the transition interval into states equivalent to the bright and dim states of the forward transition.
We take all segments of the time interval we define as the during transition interval (OBSIDs 4133010255,6), sort by some reasonable property, and compare the mean power spectra of the two halves of this sorting.
We test sorting by count rate in: various energy bands; hardness; and variability in each of the QPO and low-frequency bands. None of these separates the broadband and the QPO variability\footnote{of course, when sorting by variability, the power in the sorted band is different between the two halves, but the opposite band does not noticeably differ.}.
Therefore, either the reverse transition is truly different, in that the broadband and QPO variability are contemporaneous, or they alternate too quickly to be distinguished with the available sensitivity.

\section{Interpretation and discussion}
\label{sec:discussion}

\subsection{The Flip-Flop behaviour} \label{sec:WTFF}

One surprising property of the flip-flops considered here is how short in duration the two states can be (Figure~\ref{fig:dynamic1orbit}). This image illustrates that the source can switch from one state to the other in only a few seconds. However fast a change of state can be, the source is also observed to remain in a given state for at least thousands of seconds (limited by the length of a \nicer\ snapshot).

There are thus two clear states that the source can select from. These two states harbour major differences, yet striking similarities. The most obvious difference is in the (soft) count rate that allows us to separate them, although flux changes are barely visible at other wavelengths \citep[e.g.][figure 2d]{2023MNRAS.521.3570Y}. Moreover, both states are extremely similar in spectral shape, which points towards a negligible change in any parameter involved: accretion rate, disk and coronal densities/temperature/sizes, etc. There is thus a continuous evolution of parameters that leads to a major -- even discontinuous -- change in the behaviour: an instability?

The radiation pressure in the accretion flow could be among these parameters, as we expect it to evolve continuously with luminosity but its impact could become major above a certain value \citep[e.g.][]{1974ApJ...187L...1L, 1997ApJ...479L.145B, 2000ApJ...542L..33J}. One can imagine that the radiation pressure would become dominant in the bright state, having a major impact despite the rather small change in luminosity ($\leq 10-15\%$). As a result, the bright state would be both brighter and less stable than the dim state, which seems to be the case. Radiation pressure thus appears as the perfect candidate.

We perform some analytical estimates in Appendix \ref{sec:PhysicalState} and find that the radiation pressure should dominate by more than one order of magnitude throughout both the dim and the bright state. One can, of course, imagine that the instability requires a certain threshold $P_{\text{rad}} / P_{\text{gas}}$ higher that $1$ to be triggered, but this is beyond the scope of the present paper. Likewise, the impact of magnetic pressure is expected to be important on the stability \citep[e.g.][]{2009ApJ...697...16O,2016MNRAS.459.4397S,2022ApJ...939...31M, 2023Natur.615...45V}, which could help stabilize the bright state, or lower the impact of magnetic pressure on the dim state. Note that a significant change in the magnetic pressure should translate into changes in the emission properties of the corona. Regardless, the impact of radiation (and magnetic) pressure on the accretion flow and the flip-flops is still very much an open question.

However, while the impact of radiation pressure seems plausible, it fails to explain the striking changes in the variability pattern. During the dim state, the variability is dominated by the low-frequency broad-band noise (BBN) below $2-3$\,Hz, and there are no apparent QPO features. In turn, there is a strong QPO around $5-6$\,Hz throughout the bright state, apparently stripped of any significant BBN. While it seems at first that the flux is the trigger for the state of the system, both may instead share a common origin.

\subsection{The QPO--BBN dichotomy}

In addition to the flip-flop between the two fluxes, there is a clear dichotomy between the two states in their timing properties: the bright state shows a clear $5-6$\,Hz QPO, while the dim state only harbours a strong BBN component below $2-3$\,Hz.

It is usually considered that QPOs and BBN are two different, unrelated, features (see however the work of the vKompth group, e.g., \citealt{2020MNRAS.492.1399K}, \citealt{2022MNRAS.515.2099B}). The flip-flops observed in this study are challenging this view because, during the flip-flops, one always observes one or the other. Moreover, when sorted by hardness (see middle panels of Figures~\ref{fig:dynamic_psds} and \ref{fig:dynamic_psds_grey}), there is a hint that the QPO emerges from the BBN \footnote{Note that a link between the BBN and a (type B) QPO has been argued before \citep[see, e.g.,][]{2012MNRAS.427..595M}.}.

While this view is dependent on a sorting method that is purely empirical, the antagonism between the QPO and the BBN is undeniable in these data. This behaviour suggests that the QPO and the BBN are closely related phenomena, representing two different configurations (organizations?) of the same system.
In this view, the system would either harbour a QPO or a BBN, depending on its soft count rate, or, more likely, depending on one (or many) parameters also impacting the soft count rate. However, as discussed in section \ref{sec:DPS}, states with strong QPO and BBN are also present, as well as states with neither of them. All possible configurations are possible, and we must turn to QPO models to investigate further.

\subsection{QPO models} \label{sec:QPOmodels}

In this section we discuss the impact of our results on models for QPOs. One should keep in mind that some of these QPO models are aimed at explaining only the behaviour of type C QPOs. However, it is also important to remember that the type C QPO disappears during the hard state in this outburst, meaning that many of the arguments below are also relevant to these observations. Moreover, due to the important similarities between type B and type C QPOs it is interesting (and potentially enlightening) to seek an agreement of models to either type. Finally, there have been instances of flip-flops where type A and/or C QPOs are involved \citep[e.g.][]{2020A&A...641A.101B,2022ApJ...938..108L}, making the following discussion relevant for any types of QPOs.

There are a few candidate models to explain the behaviour of low-frequency quasi-periodic oscillations around X-ray binaries. Below, we discuss the implications of the results related to flip-flop events for what we believe are the most commonly used models in the literature. For a (much) more detailed discussion of the different models and their ability to explain the broader picture, we refer the reader to recent reviews \citep{2019NewAR..8501524I, 2022hxga.book...58D}. Note that we will not discuss the so-called relativistic precession model \citep[hereafter RPM][]{1999PhRvL..82...17S, 1999ApJ...524L..63S} because it does not make assumptions about the actual mechanism producing the QPO; which makes it impossible to assess in our study.

In the following sub-section, we investigate what could cause the change in the variability pattern, as well as the constraints it implies on the different QPO models. Because these models are all related to the state of the corona / hot flow, they do not provide an explanation for the change in soft count rate between the dim and the bright state. It is also important to note that all models considered in this paper are able to establish a configuration with(out) QPOs on timescales that are of the order of the local thermal timescale, i.e., shorter than transitions between flip-flops. Moreover, they can all in theory maintain such configuration for any duration. Therefore, the time scales involved do not directly rule out any of the QPO models.

\subsubsection{Oscillating hot corona responding to an outer disk} \label{sec:selforg}

Many of the QPO models invoke an intrinsically oscillating corona.
This corona is either radially extended as in \citet{2000astro.ph..1391P} or \citet{2010MNRAS.404..738C}, or with a less defined shape ({\citealt{2022MNRAS.515.2099B}, see also section 2 in \citealt{2020MNRAS.492.1399K}). Note that the latter has been much more thoroughly studied and compared to observations, explaining multiple observed features of the power-spectrum \citep[e.g.][]{2022MNRAS.513.4196G, 2024MNRAS.527.5638Z}, although it lacks a QPO mechanism.

These models have major characteristics in common. Namely, they rely on an external variability pattern that would 'enter' the corona, and then calculate the response of the corona to the given variability pattern. In short, the corona is assumed to act as a filter.
These models could thus explain the flip-flops observed here, as long as one can either explain that (1) the variability pattern changes during the flip-flops, or (2) the corona's response to a similar pattern changes drastically without a significant change in the coronal properties (such as its temperature or feedback fraction).

\begin{itemize}
    \item Let us first consider a change in the variability pattern. In all the models considered in this section, the variability is assumed to be communicated from the outer, optically thick, geometrically thin, cold, and turbulent accretion flow (i.e., a \citealt{1973A&A....24..337S} type of disk), through the mechanism of propagating fluctuations \citep{1997MNRAS.292..679L, 2001MNRAS.327..799K, 2005MNRAS.359..345U}. This mechanism, in which variability produced at different radii propagates inwards and combines together, naturally produces BBN, and the shape of the resultant noise is at least somewhat agnostic about the exact nature of the input variability \citep{2021MNRAS.504..469T}. Given that this process combines variability from a wide radial (and corresponding wide frequency) range, it is difficult to imagine how this could lead to a sharp feature such as a QPO. However, \citet{2023MNRAS.525.2287T} found evidence that radial epicyclic behaviour could lead to a strongly peaked feature in the power spectrum of the local accretion rate, although whether this could produce a QPO within the corona is completely untested. Even if the BBN and QPO could both originate from the outer disk in the manner described, we would expect them to be present at all times, and we therefore still require a mechanism whereby they can each be switched on and off.
    
    \item Let us now consider the second possibility: a change in the coronal properties. Unless there is major fine-tuning, any significant change of the coronal properties will have a significant impact on its spectral energy distribution. One thus needs a rather continuous change of the coronal properties to lead to a major change in its behaviour. The models discussed here need oscillation of the corona itself to have the QPO. The organisation of the oscillation of the corona around its steady-state solution would require a peculiar configuration that is achieved (resp. not-achieved) during the bright (resp. dim) state. In the case of vKompth \citep{2022MNRAS.515.2099B}, this is hard to imagine because the corona self-organizes using inverse-Compton processes that would only significantly vary due to a sudden change in the coronal properties. In the case of a radially extended corona \citep{2000astro.ph..1391P, 2010MNRAS.404..738C}, the oscillation of the corona requires the propagation of sound waves in the accretion flow. The propagation of these waves is possible as long the accretion speed $u_r$ in a hot-flow is below the sound speed $c_s$, i.e. as long as the flow is sub-sonic (over its entire radial extent). However, it has recently been argued that the accretion speed in a hot-flow during the brightest hard-states is expected to be around the sound speed, i.e. $|u_r| = c_s$, see \citet{2021ApJ...906..106M} and references therein. One can thus imagine that the speed is slightly sub-sonic in the dim state $u_r \lesssim c_s$ and slightly super-sonic $u_r \gtrsim c_s$ in the bright state; the transition happening at $u_r = c_s$. In this scenario, the ideas of both \citet{2000astro.ph..1391P} and \citet{2010MNRAS.404..738C} would align with the (dis)appearance of the QPO during the flip-flops. Note that achieving such a rapid accretion speed only needs to occur once across the entire radial span to disrupt the flow's self-organization.
\end{itemize}

\subsubsection{The impact of the jets?}

Some recent studies have argued that there is a strong connection between type B QPOs and the jets, as evidenced by numerous observations \citep{2019NewAR..8501524I}. These investigations suggest that the presence and characteristics of jets play a significant role in the formation and behaviour of these QPOs.

There is no leading candidate for type B QPOs and their link with jets. One recently proposed explanation imagines a framework where the QPO would be due to an instability triggered in the jets at a certain altitude \citep{2022A&A...660A..66F}, with a frequency linked to the Kepler frequency of the magnetic sheath at that altitude. The disturbance would propagate upstream (i.e., towards the disk) through Alfvén waves. Similarly to the previously discussed models, this model requires that (1) the instability is triggered and (2) it is able to propagate up to the corona. In the former case, the instability needs to be triggered in the bright state but not in the dim state. However, this instability remains unknown and its possibility thus cannot be addressed. In the latter case, when the QPO is absent, the upstream communication needs to be cut somehow to prevent the instability to reach the corona.

Because the idea proposed is still in its early stages it remains impossible to assess either case. However, it does possess a unique property: it requires the existence of jets to have a QPO. This is especially interesting given that flip-flops observed in a slightly softer state do not harbour QPOs \citep{2022MNRAS.513.4308L}. In other words, the flip-flops phenomenon remains the same in both cases, but the QPO can only be present when the medium generating it (the jets) exists. While this idea is promising in theory, it currently lacks numerical support and requires further investigation to establish its viability. Moreover, it seems that the compact jets have already disappeared when the flip-flops are observed in this source \citep{2021ATel14336....1T}.

\subsubsection{Lense-Thirring solid-body precession}

There are currently many strong arguments that favour geometrical models, in particular the Lense-Thirring solid-body precession model (\citealt{2009MNRAS.397L.101I}, see however \citealt{2021ApJ...906..106M}). This model assumes a misalignment between the disk and the black hole spin that provokes a frame-dragging effect. In the right conditions, this effect fuels the self-organization of the hot-flow into a precessing solid-body, comparable to a coin tossed on a flat surface before it settles. Note that despite the similarities in the name, this model is different from the RPM \citep{2018MNRAS.473..431M}.

Similarly to the aforementioned models, solid-body precession requires the self-organization of the hot flow through sound (bending) waves. However, it faces an additional challenge: the variability pattern arriving from the cold accretion disk is (by assumption) irrelevant. Instead, this model relies entirely on the properties of the hot flow (or corona), and the only explanation for the flip-flops is thus a change in the behaviour of the hot flow itself.

In this framework, when a QPO is present the hot-flow is oscillating as a solid-body at the QPO frequency. When the QPO is absent, the hot-flow is either aligned with the black hole spin or with the outer disk; either way, it is not precessing\footnote{A third possibility is that different annuli in the hot flow are precessing at different rates, but this configuration has not been theorized to our knowledge.}. In the case considered here, the presence/absence of the QPO implies a significant change in the accretion structure. Remember, however, that the spectral energy distribution of the hot flow remains unchanged during the flip-flops; as a result, the hot-flow cannot undergo significant changes in characteristics that would affect its spectral properties (temperatures, densities, aspect ratio, etc.).

In the Lense-Thirring solid-body precession theoretical framework, there are four possible configurations for the hot flow \citep{1983MNRAS.202.1181P,2009MNRAS.397L.101I,2012ApJ...757L..24N}. These four possibilities depend on (1) the ratio of the Lense-Thirring $G_{LT}$ to the viscous torque $G_\nu$ and (2) the ratio of disk scale height $\epsilon = H/R$ to viscosity $\alpha$, with each of the two ratios being independently compared to the value 1. The first condition can also be expressed in terms of $\epsilon$ and $\alpha$ \citep{2021ApJ...906..106M}, as
$$\frac{G_{LT}}{G_{\nu}} \simeq \frac{4}{3} \frac{a | \mathrm{sin} (\theta) |}{\alpha \epsilon}r^{-3/2}$$
where $a$ represents the black hole spin, $\theta$ denotes the misalignment angle between the black hole spin axis and the normal to the disk plane, and $r$ indicates the radius in units of gravitational radius. Since neither $r$ (i.e., the size of the hot-flow), $a$, nor $\theta$ are expected to change between the dim and the bright state, changes are only possible in $\alpha$ and $\epsilon$. Among the four configurations, only one harbours a QPO\footnote{Note that \citet{2021ApJ...906..106M} argued that neither condition is fulfilled in the states considered in this paper, which would rule out any possible QPO due to Lense-Thirring solid-body precession. However, for the sake of argument, we will set aside these concerns and assume that a QPO is indeed possible under these assumptions.}, when $G_{LT} \gg G_{\nu}$ and $\epsilon \gg \alpha$. This then corresponds to our bright state. However, in the dim state, no QPO is observed, and one of these conditions must cease to be fulfilled due to a slight change in either $\alpha$, $\epsilon$, or both. This issue remains to be investigated, especially since neither $\alpha$ nor $\epsilon$ are known in the inner regions of the accretion flow. However, it is difficult to imagine that a small change in either $\alpha$ or $\epsilon$ could lead to a significant change in the two conditions presented above. 

Another possible explanation for the transition from the dim to the bright state could be due to bending waves. Similar to the models discussed earlier, Lense-Thirring solid-body precession relies on the radial propagation of bending waves in the flow. However, as previously mentioned, it is anticipated that the accretion speed in the hot flow will be around the sound speed, i.e., $|u_r| = c_s$. In such a scenario, it is possible that the hot flow always remains in the configuration prone to solid-body precession ($G_{LT} \gg G_{\nu}$ and $\epsilon \gg \alpha$), with a subsonic accretion speed in the dim state $u_r \lesssim c_s$ and slightly supersonic $u_r \gtrsim c_s$ in the bright state. It is worth mentioning, however, that Lense-Thirring solid-body precession has not been studied in such regimes (see \citealt{2021ApJ...906..106M} for a discussion). We consider this to be an open question.

\section{Conclusion}
\label{sec:conclusion}

In this paper, we examine the 2021 outburst from \gx, focusing particularly on the behaviour around the hard to soft state transition, around MJD 59300. During this period, multiple flip-flops were observed over approximately three days. These flip-flops are when the source alternates between two distinct states: a bright state and a dim state. Both states exhibit very similar spectral energy distributions, differing only slightly in the black body component seen in the soft X-ray, with around a 10\% change in flux or temperature.
We show that flip-flops can occur on very short timescales, with almost 50 state changes within 1200\,s.
However, both states can also be stable over longer timescales, with entire snapshots -- greater than 1000\,s -- observed in either state, equating to millions of inner orbits. The physical cause of these two states remains a mystery beyond the scope of this study.

The bright state features a 5-6\,Hz type B Quasi-Periodic Oscillation (QPO), while the dim state is characterized by strong broadband noise (BBN). These states are clearly distinguishable, as illustrated by the dynamical power spectrum sorted by soft count rate.
We believe that the different sorting methods shown in this work (using either the count rate or the softness) demonstrate
a more continuous evolution of the physical state of the disk. While there is no justification (yet) as to why/how the variability should evolve with the count rate or softness, it is quite interesting that both these sorting methods provide much smoother variability patterns than time. It would be interesting to see the impact of such sorting method on other sources and outbursts such as the ones shown by \citealt[][MAXI J1820+070, 2018 outburst]{2020ApJ...891L..29H} or \citealt[][MAXI J1348–630, 2019 outburst]{2022ApJ...938..108L}.

Furthermore, we demonstrate that the Type-B QPO frequency increases with the count rate and/or hardness: while the frequency range is small, the frequency is definitively not constant, so some part of the mechanism which produces this QPO must couple with the effects that govern the spectral properties.

That the intermediate states merge smoothly with the prior harder and following softer states suggests that the flip-flops mark a `true' phase transition, whereas evolution outside this is largely continuous and therefore some observational states may be artificial distinctions rather than physical divisions.

Finally, we argue that the flip-flops presented in this paper provide significant insights into the QPO mechanism. Given that the corona (or hot flow) appears to have similar emission properties in both states, the type B QPO must turn on and off without significant changes in the corona's physical properties. One plausible explanation is that the corona’s accretion speed $|u_r|$ is close to its sound speed $c_s$. Since many QPO models require the propagation of radial waves, a state with $|u_r| \lesssim c_s$ could sustain QPOs, while a state with $|u_r| > c_s$ (at any radius in the hot flow) could not. In this interpretation, normal modes in the corona \citep{2000astro.ph..1391P, 2010MNRAS.404..738C} as well as Lense-Thirring solid-body precession \citep{2009MNRAS.397L.101I} could be consistent with this behaviour. It is challenging to see how other models, such as the RPM \citep{1999PhRvL..82...17S, 1999ApJ...524L..63S} or vKompth \citep{2022MNRAS.515.2099B}, could explain this behaviour. We also consider models linked to the jets \citep[e.g.,][]{2022A&A...660A..66F}, but it is difficult to explain the phenomenon since the jets seem to disappear before the flip-flops (Tremou et al., in prep.).

Despite these insights, the underlying mechanism for the flip-flops remains elusive. Specifically, the explanation for the disappearance of the QPO via $|u_r| > c_s$ does not account for the observed flip-flops. Our findings highlight the complexity of these phenomena and underscore the need for further theoretical and observational studies to unravel the mysteries of both QPO and flip-flop behaviours.

\section*{Acknowledgements}

DJKB thanks Erin Kara for useful discussions.
GM acknowledges financial support from the Polish National Science Center grant 2023/48/Q/ST9/00138 and from the Academy of Finland grant 355672.
SGDT acknowledges support under STFC Grant ST/X001113/1 and previously under an STFC studentship.
VLB thanks the European Research Council (ERC) for support under the European Union's Horizon 2020 research and innovation program (grant No. 834203) and the Department of Astronomy at the University of Maryland, College Park.

\section*{Data Availability}

The data analysed in this article are available from HEASARC.



\bibliographystyle{mnras}
\bibliography{qpo_properties}



\newpage 

\appendix

\section{Physical state of the accretion flow} \label{sec:PhysicalState}

In this section, we perform estimates of the physical state of the accretion flow in the most model-independent way. Note that, as argued in introduction, we use the ambiguous term \textit{corona} to discuss the zone that emits the X-ray emission. This zone could either be an inner hot accretion flow, the base of the jets, or a more spherical-like geometry.

In this entire section, we assume a generic black hole mass of $M=10 \, M_{\mathrm{sun}}$ \citep[see however][]{2017ApJ...846..132H} and a spin of $a=0.94$ \citep{2016ApJ...821L...6P, 2019MNRAS.484.1972J}, i.e., an inner-most stable circular orbit $R_{\mathrm{ISCO}} \simeq 2 \, GM/c^2$, where $G$ is the gravitational constant and $c$ the speed of light in vacuum. All the equations in this section can be found in either \citet{1996ApJ...465..312E}, \citet{2002apa..book.....F} and more recently \citet{2010A&A...522A..38P,2018A&A...615A..57M}.

The X-ray luminosity of the source has been measured around $L \gtrsim 20 \% \, L_\mathrm{Edd}$ \citep[Epochs 3-7]{2023ApJ...950....5L}, where $L_\mathrm{Edd}$ is the Eddington luminosity. Although approaching the slim disk state \citep{1988ApJ...332..646A}, the disk is still expected to be in a standard SS73 state. Assuming a typical radiative efficiency of $10\%$ \citep{2014ARA&A..52..529Y}, one can estimate an expected accretion rate $\dot{M} \gtrsim 2 \, L_{\mathrm{Edd}}/c^2$. In turn, the disk temperature can be estimated by fitting the black-body component in the spectra. Assuming no colour-correction factor, \citet{2023MNRAS.521.3570Y} found $k_B T \lesssim 1 \, \mathrm{keV}$ ($k_B$ the Boltzmann constant) in the states considered in the present study.

Let us here assume that the black-body component is produced by a typical electro-neutral, one-temperature, cold disk. All quantities are expressed at a given radius $R$. In such case, one can write the gaseous pressure $P_{\mathrm{gas}}$ as a function of the disk density $n$, the Boltzmann constant $k_B$, and the disk central temperature $T_c$ as 
\begin{equation}
    P_{\mathrm{gas}} \simeq 2 n k_B \, T_c \simeq 2 n k_B \, \tau^{1/4} \, T \label{eq:pgas}
\end{equation}
where $T$ is the disk surface (i.e. measured) temperature, and $\tau$ the vertical optical depth.
In turn, the radiative pressure $P_{\mathrm{rad}}$ can be expressed as a function of the locally dissipated cooling rate $q_-$ and the disk half-thickness $H$, such that
\begin{equation}
    P_{\mathrm{rad}} \simeq q_- \frac{H}{2 c} \tau \label{eq:prad}
\end{equation}
In the cold disk considered, all the local accretion power is usually considered to be radiated away. In other words,
\begin{equation}
    q_- = q_+ = \frac{G M \dot{M}}{8 \pi H R^3} \label{eq:q-}
\end{equation}

In the temperatures and densities considered in this study we lie in the typical Thomson regime $\tau = \rho H \sigma_T / m_p$, with $\rho$ the local density, $\sigma_T$ the Stefan-Boltzmann constant, and $m_p$ the proton mass. Moreover, we argued above that $\dot{M} c^2 \gtrsim L_{\mathrm{Edd}}$ in the state considered. Combining equations (\ref{eq:prad}) and (\ref{eq:q-}) thus leads to the following estimate for the radiative pressure
\begin{equation}
    P_{\mathrm{rad}} \gtrsim \frac{n m_p c^2}{4} \, \frac{H}{R} \, \left( \frac{1}{R / R_g} \right)^2 \, \left( \frac{\dot{M} c^2}{1 \, L_{\mathrm{Edd}}} \right)
\end{equation}
which then leads to
\begin{equation}
    \frac{P_{\mathrm{rad}}}{P_{\mathrm{gas}}} \simeq \frac{1}{4 \tau^{1/4}} \, \frac{m_p c^2}{k_B T} \, \frac{H}{R} \, \left( \frac{1}{R / R_g} \right)^2 \, \left( \frac{\dot{M} c^2}{1 \, L_{\mathrm{Edd}}} \right)
\end{equation}

We assume that the thermal disk extends almost all the way down to $R_{\mathrm{ISCO}}$, say $R \sim 2 \, R_{\mathrm{ISCO}} \sim 4 \, R_g$, which is consistent with reflection modelling in these epochs ($R \sim 1-3 \, R_{\mathrm{ISCO}}$, \citealt{2023ApJ...950....5L}).
We expect the disk to still be relatively thin at such high accretion rate, i.e. with an aspect ratio $H/R \gtrsim 10^{-2}$. The estimate for $\tau$ is assumption-dependent, but with realistic assumptions of $H/R = 10^{-2}$ and $n \lesssim 10^{22}$ at $R=4\,R_g$, one obtains $\tau \lesssim 400$. These likely assumptions lead to 
\begin{equation}
    \frac{P_{\mathrm{rad}}}{P_{\mathrm{gas}}} \gtrsim 30 \, \left( \frac{\tau}{400} \right)^{-1} \,\left( \frac{k_B T}{1\,\mathrm{keV}} \right)^{-1} \, \left( \frac{H/R}{10^{-2}} \right) \, \left( \frac{R}{4 \, R_g} \right)^{-2} \, \left( \frac{\dot{M} c^2}{1 \, L_{\mathrm{Edd}}} \right)
\end{equation}
Note that the JED-SAD model \citep{2018A&A...617A..46M} with the best parameters to reproduce the spectral shape ($a = 0.94$, $R_J \simeq 6 \, R_g$, and $\dot{M}_{\mathrm{in}} \simeq 2 \, L_{\mathrm{Edd}}/c^2$) leads to $P_{\mathrm{rad}} \gtrsim 100 \, P_{\mathrm{gas}}$ in the thermal disk, i.e. a comparable value.

This estimate lies in a regime where the radiation pressure strongly dominates over the gaseous pressure.

\section{Fitting power spectra}
\label{app:psfitting}
\label{sec:psdfitting}

Fitting averaged power spectra is a standard task; however, fitting individual periodograms is less common so we include details here.

Fitting power spectra is often done by averaging over many light curve segments, allowing errors to be treated as approximately Gaussian.
However, when considering changes in the PSD which occur on shorter time scales than the total length of the light curve segments to be averaged, this is not possible.
Instead, we may fit periodograms individually, accounting for the non-Gaussian distribution of periodogram points directly.

For many random processes, the Fourier transform of a segment may be treated as a Gaussian distribution, with each component independent.
This is shown to be asymptotically accurate for Poisson noise with a constant mean (i.e. an intrinsically constant signal) in \citet{1983ApJ...266..160L}.
Fitting since then sometimes implicitly assumes that this is still the case for more general processes \citep[e.g.[]{}.
We do not guarantee that our light curves are thus distributed and instead check after fitting that the observed spectral powers follow this distribution well enough.

In this case, the likelihood function for each frequency in the periodogram is
$$P(d|m) = me^{-\frac{d}{m}}$$
and the overall likelihood is the product of these terms across the frequencies  and periodograms used (in practice, this is calculated in log space).
The model values may then be a function of both the frequency and any other properties of the light curve segment used to produce each periodogram (e.g. its hardness).

We then use this likelihood in a standard Bayesian formalism; a description of the priors chosen is given at the relevant points in the text.

As is common, our models for power spectra consist of sums of Lorentzian functions.
There are several common parametrisations of a Lorentzian. We choose
$$L(\nu;(N,\nu_{\rm c},\nu_{\rm w})) = \frac{N}{\pi\nu_{\rm w}\left(1+\left(\frac{\nu-\nu_{\rm c}}{\nu_{\rm w}}\right)^2\right)} $$
This choice of normalisation gives $\int_{-\infty}^{\infty}Ld\nu=N$; in cases where significant power is present outside the fitted frequency range $[\nu_0,\nu_1]$, we instead choose $\int_{\nu_0}^{\nu_1}Ld\nu=N$ for stability when fitting. Explicitly,
$$L(\nu;(N,\nu_{\rm c},\nu_{\rm w})) = \frac{1}{\tan^{-1}\left(\frac{\nu_1-\nu_c}{\nu_w}\right)-\tan^{-1}\left(\frac{\nu_0-\nu_c}{\nu_w}\right)}
\frac{N}{\nu_{\rm w}\left(1+\left(\frac{\nu-\nu_{\rm c}}{\nu_{\rm w}}\right)^2\right)} $$

The use of Lorentzian functions is motivated theoretically by their being the power spectrum of a damped random walk.
Empirically, this often gives a good fit to power spectra using only a few components.
To maintain a close link to a simple ACF, we symmetrise the Lorentzian function by adding a component with $\nu\to-\nu$, i.e.
$$L_\pm(\nu;(N,\nu_{\rm c},\nu_{\rm w})) = \frac{N}{\pi\nu_{\rm w}\left(1+\left(\frac{\nu-\nu_{\rm c}}{\nu_{\rm w}}\right)^2\right)}  +  \frac{N}{\pi\nu_{\rm w}\left(1+\left(\frac{\nu+\nu_{\rm c}}{\nu_{\rm w}}\right)^2\right)} $$
This is slightly cleaner than ignoring the implicit symmetrisation\footnote{i.e. just considering the positive frequencies when fitting, $P(\nu)=H(\nu)\frac{N}{\pi\nu_{\rm w}\left(1+\left(\frac{\nu-\nu_{\rm c}}{\nu_{\rm w}}\right)^2\right)}  +  H(-\nu)\frac{N}{\pi\nu_{\rm w}\left(1+\left(\frac{\nu+\nu_{\rm c}}{\nu_{\rm w}}\right)^2\right)} $, where $H$ is the Heaviside step function.} but equivalent in each of the limits $\nu_{\rm c}\ll\nu_{\rm w}$ and $\nu_{\rm c}\gg\nu_{\rm w}$

The changes of power spectral parameters relative to spectral emission changes do not have full theoretical descriptions.
Instead, for small enough changes in spectral parameters, we may expand the relation as a power series.
Denoting the spectral parameter $p_{\rm s}$, we then have
$$ P(\nu) = \sum_i L(\nu;\theta_i)$$
with
$$ \theta_i = \theta_{i,0} + (p_{\rm s} - p_{{\rm s},0})\theta_{i,1} + (p_{\rm s} - p_{{\rm s},0})^2\theta_{i,2} + ... $$

where $p_{{\rm s},0}$ is the value of spectral parameter about which the expansion is taken (we use the median of the values from the segments used). In each case, $\theta_{i,j}$ is a vector of the three parameters of a single Lorentzian.
We truncate the power series at the first term $j$ where $\theta_{i,j}$ is consistent with 0.

\subsection{Binning and windowing effects}

The light curve measurements are bin averages rather than point measurements. This is equivalent to measuring the intrinsic light curve convolved with a rectangle function with width equal to the bin size.
The effect of this on the power spectrum is that of multiplication by the square of the Fourier transform of this rectangle function, 
$$P_{\rm Meas}(\nu) = P_{\rm int}(\nu) \left( \frac{\sin \pi\nu\Delta t}{\pi \nu \Delta t} \right)^2$$
Since we measure frequencies much lower than the Nyquist frequency, this is a small effect. For the highest frequency shown (30\,Hz), $P_{\rm Meas}/P_{\rm int} = 0.9985$. Strictly, there is also a contribution to the measured power from frequencies above the Nyquist frequency, but these are expected to be small and further suppressed by the $(\nu\Delta t)^2$ term. We ignore these effects in calculations for computational efficiency.

Additionally, the use of short finite segments means that power at low Fourier frequencies is transferred to the frequencies which are represented in the measured power spectrum (leakage).
We do not attempt to correct for this in images or models but note the likely effect where relevant in the text.
This typically adds power with a spectral shape which decreases with frequency.

\begin{table*}
    \centering
\begin{tabular}{lrrrr}
\hline
\multirow{2}{1.5cm}{Parameter} & \multicolumn{4}{c}{Sorting property} \\
& Hardness ($4-10$/$2-4$\,keV) & Time (s) & Hard rate ($4-10$\,keV cts/s) & Soft rate ($2-4$\,keV cts/s) \\
\hline
QPO norm                          & $ 4.28   _{- 0.23   }^{+ 0.183  } \times 10^{-4}$ & $ 4.94   _{- 0.262  }^{+ 0.266  } \times 10^{-4}$ & $ 3.97   _{- 0.177  }^{+ 0.167  } \times 10^{-4}$ & $ 4.4    _{- 0.22   }^{+ 0.219  } \times 10^{-4}$ \\
$\Delta\nu_{\rm QPO}$            & $ 0.412  _{- 0.0237 }^{+ 0.0225 }$& $ 0.46   _{- 0.0263 }^{+ 0.0266 }$& $ 0.351  _{- 0.0173 }^{+ 0.0201 }$& $ 0.417  _{- 0.0222 }^{+ 0.0317 }$ \\
$\nu_{\rm QPO}$                  & $ 5.12   _{- 0.0126 }^{+ 0.0143 }$& $ 5.15   _{- 0.0174 }^{+ 0.0201 }$& $ 5.06   _{- 0.0128 }^{+ 0.0116 }$& $ 5.03   _{- 0.0166 }^{+ 0.0144 }$ \\
Noise norm                        & $ 1.12   _{- 0.302  }^{+ 0.307  } \times 10^{-4}$ & $ 4.5    _{- 2.84   }^{+ 3.25   } \times 10^{-5}$ & $ 1.16   _{- 0.248  }^{+ 0.267  } \times 10^{-4}$ & $ 6.09   _{- 2.98   }^{+ 2.51   } \times 10^{-5}$ \\
$\Delta\nu_{\rm Noise}$          & $ 25.6   _{- 19     }^{+ 17.7   }$& $ 28.5   _{- 19.9   }^{+ 15.4   }$& $ 23.2   _{- 17.5   }^{+ 20.2   }$& $ 26.4   _{- 20.9   }^{+ 15.3   }$ \\
\hline
$\nabla$ QPO norm                & $ 0.0112 _{- 0.00551}^{+ 0.00442}$ & $ 1.41   _{- 7.25   }^{+ 8.51   } \times 10^{-10}$ & $ 8.53   _{- 2.21   }^{+ 2.55   } \times 10^{-6}$ & $ 2.29   _{- 0.848  }^{+ 0.838  } \times 10^{-6}$ \\
$\nabla\Delta\nu_{\rm QPO}$      & $ 24.8   _{- 6.75   }^{+ 5.41   }$ & $-2.67   _{- 0.759  }^{+ 1.01   } \times 10^{-6}$ & $ 9.66   _{- 2.72   }^{+ 2.64   } \times 10^{-3}$ & $ 9.19   _{- 10.8   }^{+ 9.7    } \times 10^{-4}$ \\
$\nabla\nu_{\rm QPO}$            & $ 49.7   _{- 3.68   }^{+ 3.26   }$ & $-6.71   _{- 0.434  }^{+ 0.507  } \times 10^{-6}$& $ 0.0322 _{- 0.00152}^{+ 0.002  }$ & $ 3.61   _{- 0.591  }^{+ 0.652  } \times 10^{-3}$ \\
$\nabla$ Noise norm              & $ 0.0109 _{- 0.00683}^{+ 0.00661}$ & $-3.56   _{- 0.785  }^{+ 0.955  } \times 10^{-9}$ & $ 3.31   _{- 3.22   }^{+ 3.49   } \times 10^{-6}$ & $-2.54   _{- 1.12   }^{+ 1.03   } \times 10^{-6}$ \\
$\nabla \Delta\nu_{\rm Noise}$    & $-9.56   _{- 438    }^{+ 399    } \times 10^{8}$ & $ 7.41   _{- 3.74e+03}^{+ 3.76e+03} \times 10^{-4}$ & $-1.64   _{- 124    }^{+ 135    } \times 10^{3}$& $-33.7   _{- 288    }^{+ 266    }$ \\
\hline
$\nabla^2$ QPO norm              & $ 0.534  _{- 0.412  }^{+ 0.363  }$ & $-3.27   _{- 1.79   }^{+ 2.21   } \times 10^{-14}$ & $ 3.91   _{- 1.58   }^{+ 1.64   } \times 10^{-7}$ & $ 4.12   _{- 11.1   }^{+ 11.4   } \times 10^{-9}$ \\
$\nabla^2\Delta\nu_{\rm QPO}$    & $ 234    _{- 352    }^{+ 445    }$ & $-2.18   _{- 2.03   }^{+ 2.33   } \times 10^{-11}$ & $ 1.57   _{- 1.45   }^{+ 1.89   } \times 10^{-4}$ & $ 1.67   _{- 1.33   }^{+ 1.49   } \times 10^{-5}$ \\
$\nabla^2\nu_{\rm QPO}$          & $ 725    _{- 281    }^{+ 265    }$ & $ 1.35   _{- 11.6   }^{+ 12.9   } \times 10^{-12}$ & $ 3.41   _{- 1.28   }^{+ 1.01   } \times 10^{-4}$ & $ 5.04   _{- 0.906  }^{+ 0.713  } \times 10^{-5}$ \\
$\nabla^2$ Noise norm            & $-1.1    _{- 0.652  }^{+ 0.572  }$ & $-1.64   _{- 22     }^{+ 21.6   } \times 10^{-15}$ & $-4.76   _{- 2.51   }^{+ 2.39   } \times 10^{-7}$ & $ 1.64   _{- 1.38   }^{+ 1.26   } \times 10^{-8}$ \\
$\nabla^2 \Delta\nu_{\rm Noise}$  & $ 2.88   _{- 22     }^{+ 17.1   } \times 10^{10}$ & $ 2.13   _{- 217    }^{+ 261    } \times 10^{-8}$& $-108    _{- 2.38e+03}^{+ 2.71e+03}$ & $-3.26   _{- 1.28e+03}^{+ 1.13e+03} \times 10^{-3}$ \\
\hline
$\nabla^3$ QPO norm              & $-67.8   _{- 46.7   }^{+ 50.6   }$ & $-2.39   _{- 1.43   }^{+ 1.6    } \times 10^{-19}$ & $-2.17   _{- 0.811  }^{+ 0.798  } \times 10^{-8}$ & $-2.72   _{- 1.67   }^{+ 2.66   } \times 10^{-10}$ \\
$\nabla^3\Delta\nu_{\rm QPO}$     & $-1.11   _{- 0.518  }^{+ 0.545  } \times 10^{5}$ & $-9.43   _{- 16.8   }^{+ 18.1   } \times 10^{-17}$ & $-1.75   _{- 0.75   }^{+ 0.735  } \times 10^{-5}$ & $-3.79   _{- 2.22   }^{+ 3.26   } \times 10^{-7}$ \\
$\nabla^3\nu_{\rm QPO}$           & $-9.24   _{- 2.87   }^{+ 3.85   } \times 10^{4}$ & $ 3.11   _{- 0.922  }^{+ 1.01   } \times 10^{-16}$ & $-2.14   _{- 0.619  }^{+ 0.515  } \times 10^{-5}$ & $ 2.95   _{- 1.71   }^{+ 1.77   } \times 10^{-7}$ \\
$\nabla^3$ Noise norm            & $ 5.39   _{- 61.7   }^{+ 66.6   }$ & $ 1.27   _{- 1.64   }^{+ 1.77   } \times 10^{-19}$ & $ 1.7    _{- 1.28   }^{+ 1.26   } \times 10^{-8}$ & $ 4.74   _{- 2.51   }^{+ 2.01   } \times 10^{-10}$ \\
$\nabla^3 \Delta\nu_{\rm Noise}$  & $ 4.67   _{- 44.3   }^{+ 46     } \times 10^{9}$ & $-1.59   _{- 34.2   }^{+ 34.7   } \times 10^{-13}$& $-1.52   _{- 14.7   }^{+ 15.1   }$ & $ 3.1    _{- 394    }^{+ 426    } \times 10^{-5}$ \\
\hline
    \end{tabular}
    \caption{Table of QPO parameters for model including first three derivatives of each basic parameter relative to a property of each segment. $\nabla$ indicates the derivative wrt the value of the sorting property.}
    \label{tab:qpo_pars3}
\end{table*}

\begin{table*}
    \centering
\begin{tabular}{lrrrr}
\hline
\multirow{2}{1.5cm}{Parameter} & \multicolumn{4}{c}{Sorting property} \\
& Hardness ($4-10$/$2-4$\,keV) & Time (s) & Hard rate ($4-10$\,keV cts/s) & Soft rate ($2-4$\,keV cts/s) \\
\hline
QPO norm                          & $ 4.3    _{- 0.23   }^{+ 0.231  } \times 10^{-4}$ & $ 5.07   _{- 0.237  }^{+ 0.226  } \times 10^{-4}$ & $ 4.24   _{- 0.201  }^{+ 0.213  } \times 10^{-4}$ & $ 4.26   _{- 0.218  }^{+ 0.233  } \times 10^{-4}$ \\
$\Delta\nu_{\rm QPO}$            & $ 0.413  _{- 0.0252 }^{+ 0.0267 }$& $ 0.478  _{- 0.0275 }^{+ 0.0284 }$& $ 0.379  _{- 0.0218 }^{+ 0.0226 }$& $ 0.423  _{- 0.0271 }^{+ 0.0289 }$ \\
$\nu_{\rm QPO}$                  & $ 5.12   _{- 0.0164 }^{+ 0.0168 }$& $ 5.16   _{- 0.0203 }^{+ 0.0202 }$& $ 5.07   _{- 0.0138 }^{+ 0.0133 }$& $ 5      _{- 0.0171 }^{+ 0.0174 }$ \\
Noise norm                        & $ 1.27   _{- 0.323  }^{+ 0.322  } \times 10^{-4}$ & $ 2.72   _{- 1.96   }^{+ 3.1    } \times 10^{-5}$ & $ 8.98   _{- 2.96   }^{+ 2.83   } \times 10^{-5}$ & $ 7.57   _{- 3.15   }^{+ 3.02   } \times 10^{-5}$ \\
$\Delta\nu_{\rm Noise}$          & $ 24.3   _{- 17.3   }^{+ 17.8   }$& $ 24.5   _{- 16.9   }^{+ 18     }$& $ 25     _{- 17.6   }^{+ 17.7   }$& $ 27.2   _{- 18.3   }^{+ 15.8   }$ \\
\hline
$\nabla$ QPO norm                & $ 0.0107 _{- 0.00556}^{+ 0.00563}$ & $-2.36   _{- 1.18   }^{+ 1.11   } \times 10^{-9}$ & $ 1.4    _{- 0.346  }^{+ 0.342  } \times 10^{-5}$ & $ 2.01   _{- 0.899  }^{+ 0.823  } \times 10^{-6}$ \\
$\nabla\Delta\nu_{\rm QPO}$      & $ 24     _{- 6.3    }^{+ 6.42   }$ & $-5.69   _{- 1.41   }^{+ 1.38   } \times 10^{-6}$& $ 0.0145 _{- 0.00346}^{+ 0.00344}$ & $ 1.09   _{- 1.13   }^{+ 1.06   } \times 10^{-3}$ \\
$\nabla\nu_{\rm QPO}$            & $ 49.1   _{- 3.65   }^{+ 3.65   }$ & $-8.37   _{- 0.839  }^{+ 0.83   } \times 10^{-6}$& $ 0.0335 _{- 0.00229}^{+ 0.00221}$ & $ 2.8    _{- 0.672  }^{+ 0.658  } \times 10^{-3}$ \\
$\nabla$ Noise norm              & $ 0.0116 _{- 0.00733}^{+ 0.00731}$ & $-2.27   _{- 1.49   }^{+ 1.49   } \times 10^{-9}$ & $-1.25   _{- 4.87   }^{+ 5.04   } \times 10^{-6}$ & $-2.12   _{- 1.02   }^{+ 1.05   } \times 10^{-6}$ \\
$\nabla \Delta\nu_{\rm Noise}$    & $-2.08   _{- 938    }^{+ 950    } \times 10^{9}$& $-0.0308 _{- 22.3   }^{+ 17.9   }$& $ 367    _{- 1.33e+06}^{+ 1.55e+06}$& $ 0.655  _{- 2.97e+03}^{+ 3.79e+03}$ \\
\hline
$\nabla^2$ QPO norm              & $ 0.0332 _{- 1.02   }^{+ 1.03   }$ & $-2.92   _{- 1.67   }^{+ 1.78   } \times 10^{-14}$ & $-2.26   _{- 2.74   }^{+ 2.72   } \times 10^{-7}$ & $ 2.36   _{- 2.12   }^{+ 2.06   } \times 10^{-8}$ \\
$\nabla^2\Delta\nu_{\rm QPO}$    & $-125    _{- 1.15e+03}^{+ 1.16e+03}$ & $-1.81   _{- 1.98   }^{+ 2.06   } \times 10^{-11}$ & $-4.07   _{- 2.81   }^{+ 2.74   } \times 10^{-4}$ & $ 2.06   _{- 26.8   }^{+ 26.2   } \times 10^{-6}$ \\
$\nabla^2\nu_{\rm QPO}$           & $ 1.3    _{- 0.699  }^{+ 0.697  } \times 10^{3}$ & $-1.75   _{- 13.1   }^{+ 13     } \times 10^{-12}$ & $ 1.83   _{- 1.7    }^{+ 1.7    } \times 10^{-4}$ & $ 1.19   _{- 0.208  }^{+ 0.211  } \times 10^{-4}$ \\
$\nabla^2$ Noise norm            & $-2.05   _{- 1.4    }^{+ 1.41   }$ & $ 8.72   _{- 23.1   }^{+ 22.3   } \times 10^{-15}$ & $ 8.77   _{- 37.9   }^{+ 37.3   } \times 10^{-8}$ & $-8.13   _{- 25     }^{+ 24.4   } \times 10^{-9}$ \\
$\nabla^2 \Delta\nu_{\rm Noise}$  & $ 1.8    _{- 1.08e+03}^{+ 964    } \times 10^{9}$ & $-2.69   _{- 1.6e+03}^{+ 1.11e+03} \times 10^{-7}$& $-5.07   _{- 3.38e+04}^{+ 2.54e+04}$& $ 0.0201 _{- 7.39   }^{+ 8.8    }$ \\
\hline
$\nabla^3$ QPO norm              & $-72.6   _{- 58.8   }^{+ 61.8   }$ & $ 9.1    _{- 4.27   }^{+ 4.4    } \times 10^{-19}$ & $-6.06   _{- 1.95   }^{+ 1.83   } \times 10^{-8}$ & $-9.72   _{- 19.6   }^{+ 19.4   } \times 10^{-11}$ \\
$\nabla^3\Delta\nu_{\rm QPO}$     & $-1.13   _{- 0.628  }^{+ 0.671  } \times 10^{5}$ & $ 1.22   _{- 0.49   }^{+ 0.507  } \times 10^{-15}$ & $-5.09   _{- 1.93   }^{+ 1.68   } \times 10^{-5}$ & $-3.22   _{- 3.91   }^{+ 3.21   } \times 10^{-7}$ \\
$\nabla^3\nu_{\rm QPO}$           & $-7.9    _{- 3.82   }^{+ 3.82   } \times 10^{4}$ & $ 9.86   _{- 3.03   }^{+ 3.23   } \times 10^{-16}$ & $-2.84   _{- 1.07   }^{+ 1.14   } \times 10^{-5}$ & $ 5.61   _{- 1.67   }^{+ 1.72   } \times 10^{-7}$ \\
$\nabla^3$ Noise norm            & $ 0.384  _{- 80     }^{+ 76.3   }$ & $-4.24   _{- 5.71   }^{+ 5.75   } \times 10^{-19}$ & $ 4.9    _{- 2.53   }^{+ 2.48   } \times 10^{-8}$ & $ 2.37   _{- 2.21   }^{+ 1.95   } \times 10^{-10}$ \\
$\nabla^3 \Delta\nu_{\rm Noise}$  & $ 2.23   _{- 7.65e+03}^{+ 6.77e+03} \times 10^{8}$ & $-2.84   _{- 1.84e+03}^{+ 1.49e+03} \times 10^{-13}$& $-0.161  _{- 590    }^{+ 500    }$ & $ 3.04   _{- 4.11e+03}^{+ 5.44e+03} \times 10^{-5}$ \\
\hline
$\nabla^4$ QPO norm               & $ 4.44   _{- 6.38   }^{+ 7.31   } \times 10^{3}$ & $ 7.16   _{- 2.49   }^{+ 2.45   } \times 10^{-24}$ & $ 1.98   _{- 0.725  }^{+ 0.799  } \times 10^{-9}$ & $-2.13   _{- 2.43   }^{+ 3.25   } \times 10^{-12}$ \\
$\nabla^4\Delta\nu_{\rm QPO}$     & $ 3.54   _{- 6.77   }^{+ 8.38   } \times 10^{6}$ & $ 8.13   _{- 2.91   }^{+ 2.82   } \times 10^{-21}$ & $ 1.74   _{- 0.709  }^{+ 0.758  } \times 10^{-6}$ & $ 3.15   _{- 4.12   }^{+ 4.99   } \times 10^{-9}$ \\
$\nabla^4\nu_{\rm QPO}$           & $-4.15   _{- 4.54   }^{+ 4.37   } \times 10^{6}$ & $ 4.34   _{- 1.92   }^{+ 1.98   } \times 10^{-21}$ & $ 4.32   _{- 4.51   }^{+ 4.27   } \times 10^{-7}$ & $-1.29   _{- 0.358  }^{+ 0.338  } \times 10^{-8}$ \\
$\nabla^4$ Noise norm             & $ 5.13   _{- 9.41   }^{+ 9.09   } \times 10^{3}$ & $-4.17   _{- 2.97   }^{+ 3.19   } \times 10^{-24}$ & $-1.7    _{- 0.986  }^{+ 0.999  } \times 10^{-9}$ & $ 2.48   _{- 2.8    }^{+ 2.55   } \times 10^{-12}$ \\
$\nabla^4 \Delta\nu_{\rm Noise}$  & $-9.07   _{- 8.42e+03}^{+ 8.11e+03} \times 10^{8}$ & $-5.11   _{- 2.32e+03}^{+ 1.82e+03} \times 10^{-19}$ & $ 4.18   _{- 3.92e+03}^{+ 4.77e+03} \times 10^{-3}$ & $-1.12   _{- 1.26e+03}^{+ 1.33e+03} \times 10^{-7}$ \\
\hline
    \end{tabular}
    \caption{Table of QPO parameters for model including first four derivatives of each basic parameter relative to a property of each segment. $\nabla$ indicates the derivative wrt the value of the sorting property.
    }
    \label{tab:qpo_pars4}
\end{table*}

\begin{table*}
    \centering
\begin{tabular}{lrrrr}
\hline
\multirow{2}{1.5cm}{Parameter} & \multicolumn{4}{c}{Sorting property} \\
& Hardness ($4-10$/$2-4$\,keV) & Time (s) & Hard rate ($4-10$\,keV cts/s) & Soft rate ($2-4$\,keV cts/s) \\
\hline
QPO norm                          & $ 4.34   _{- 0.247  }^{+ 0.249  } \times 10^{-4}$ & $ 5.27   _{- 0.353  }^{+ 0.315  } \times 10^{-4}$ & $ 4.09   _{- 0.214  }^{+ 0.222  } \times 10^{-4}$ & $ 4.23   _{- 0.229  }^{+ 0.236  } \times 10^{-4}$ \\
$\Delta\nu_{\rm QPO}$            & $ 0.419  _{- 0.0276 }^{+ 0.0302 }$& $ 0.476  _{- 0.0378 }^{+ 0.0369 }$& $ 0.367  _{- 0.0238 }^{+ 0.0241 }$& $ 0.413  _{- 0.0269 }^{+ 0.0282 }$ \\
$\nu_{\rm QPO}$                  & $ 5.12   _{- 0.0169 }^{+ 0.0168 }$& $ 5.16   _{- 0.0267 }^{+ 0.0274 }$& $ 5.07   _{- 0.0154 }^{+ 0.0146 }$& $ 5      _{- 0.0168 }^{+ 0.0176 }$ \\
Noise norm                        & $ 1.25   _{- 0.346  }^{+ 0.333  } \times 10^{-4}$ & $ 3.76   _{- 2.69   }^{+ 4.26   } \times 10^{-5}$ & $ 1.09   _{- 0.312  }^{+ 0.314  } \times 10^{-4}$ & $ 7.86   _{- 3.07   }^{+ 3.06   } \times 10^{-5}$ \\
$\Delta\nu_{\rm Noise}$          & $ 25.6   _{- 17.8   }^{+ 17     }$& $ 25     _{- 17.5   }^{+ 17.3   }$& $ 25.6   _{- 17.9   }^{+ 17.2   }$& $ 25.9   _{- 17.3   }^{+ 17     }$ \\
\hline
$\nabla$ QPO norm                & $ 0.0117 _{- 0.00707}^{+ 0.00697}$ & $-2.95   _{- 1.54   }^{+ 1.48   } \times 10^{-9}$ & $ 1.73   _{- 0.386  }^{+ 0.404  } \times 10^{-5}$ & $ 2.21   _{- 1.1    }^{+ 1.05   } \times 10^{-6}$ \\
$\nabla\Delta\nu_{\rm QPO}$      & $ 29     _{- 8.49   }^{+ 8.57   }$ & $-5.11   _{- 1.75   }^{+ 1.66   } \times 10^{-6}$& $ 0.0172 _{- 0.00378}^{+ 0.00401}$ & $ 2.86   _{- 1.47   }^{+ 1.32   } \times 10^{-3}$ \\
$\nabla\nu_{\rm QPO}$            & $ 52.1   _{- 5.64   }^{+ 5.6    }$ & $-8.43   _{- 1.03   }^{+ 1.06   } \times 10^{-6}$& $ 0.0337 _{- 0.0025 }^{+ 0.00256}$ & $ 3.43   _{- 0.901  }^{+ 0.914  } \times 10^{-3}$ \\
$\nabla$ Noise norm               & $ 9.91   _{- 10.3   }^{+ 10.1   } \times 10^{-3}$ & $-4.02   _{- 1.91   }^{+ 1.91   } \times 10^{-9}$ & $-5.28   _{- 5.18   }^{+ 5.37   } \times 10^{-6}$ & $-2.84   _{- 1.4    }^{+ 1.36   } \times 10^{-6}$ \\
$\nabla \Delta\nu_{\rm Noise}$    & $ 1.41   _{- 1.41e+04}^{+ 1.2e+04} \times 10^{7}$ & $-1.69   _{- 1.78e+03}^{+ 1.57e+03} \times 10^{-3}$& $ 42.8   _{- 7.55e+03}^{+ 8.42e+03}$& $-0.328  _{- 181    }^{+ 195    }$ \\
\hline
$\nabla^2$ QPO norm              & $-0.268  _{- 1.18   }^{+ 1.09   }$ & $-5.04   _{- 4.44   }^{+ 4.63   } \times 10^{-14}$ & $ 4.76   _{- 4.64   }^{+ 4.65   } \times 10^{-7}$ & $ 2.77   _{- 2.42   }^{+ 2.4    } \times 10^{-8}$ \\
$\nabla^2\Delta\nu_{\rm QPO}$    & $-675    _{- 1.36e+03}^{+ 1.33e+03}$ & $-9.83   _{- 522    }^{+ 512    } \times 10^{-13}$ & $ 1.2    _{- 4.81   }^{+ 4.76   } \times 10^{-4}$ & $ 2.12   _{- 2.97   }^{+ 3.12   } \times 10^{-5}$ \\
$\nabla^2\nu_{\rm QPO}$           & $ 1.31   _{- 0.741  }^{+ 0.767  } \times 10^{3}$ & $ 9.94   _{- 341    }^{+ 314    } \times 10^{-13}$ & $ 2.65   _{- 3.03   }^{+ 3.18   } \times 10^{-4}$ & $ 1.28   _{- 0.219  }^{+ 0.228  } \times 10^{-4}$ \\
$\nabla^2$ Noise norm            & $-1.89   _{- 1.49   }^{+ 1.49   }$ & $-2.19   _{- 5.12   }^{+ 4.51   } \times 10^{-14}$ & $-7.84   _{- 6.54   }^{+ 6.58   } \times 10^{-7}$ & $-1.41   _{- 2.8    }^{+ 2.59   } \times 10^{-8}$ \\
$\nabla^2 \Delta\nu_{\rm Noise}$  & $ 7.81   _{- 982    }^{+ 1.09e+03} \times 10^{8}$ & $-2.78   _{- 747    }^{+ 642    } \times 10^{-8}$& $ 0.561  _{- 331    }^{+ 375    }$& $ 0.0105 _{- 0.886  }^{+ 1.3    }$ \\
\hline
$\nabla^3$ QPO norm              & $-92.7   _{- 144    }^{+ 112    }$ & $ 1.12   _{- 0.685  }^{+ 0.712  } \times 10^{-18}$ & $-9.69   _{- 2.84   }^{+ 2.74   } \times 10^{-8}$ & $-2.23   _{- 4.57   }^{+ 5      } \times 10^{-10}$ \\
$\nabla^3\Delta\nu_{\rm QPO}$     & $-2.74   _{- 1.96   }^{+ 1.84   } \times 10^{5}$ & $ 7.97   _{- 7.46   }^{+ 7.69   } \times 10^{-16}$ & $-8.03   _{- 2.56   }^{+ 2.57   } \times 10^{-5}$ & $-1.58   _{- 0.744  }^{+ 0.756  } \times 10^{-6}$ \\
$\nabla^3\nu_{\rm QPO}$           & $-1.64   _{- 1.4    }^{+ 1.3    } \times 10^{5}$ & $ 1.01   _{- 0.451  }^{+ 0.47   } \times 10^{-15}$ & $-3.13   _{- 1.62   }^{+ 1.68   } \times 10^{-5}$ & $ 5.62   _{- 55.1   }^{+ 56.5   } \times 10^{-8}$ \\
$\nabla^3$ Noise norm            & $ 51.6   _{- 206    }^{+ 211    }$ & $ 6.16   _{- 9.19   }^{+ 8.7    } \times 10^{-19}$ & $ 9.13   _{- 3.44   }^{+ 3.46   } \times 10^{-8}$ & $ 6.17   _{- 5.26   }^{+ 5.06   } \times 10^{-10}$ \\
$\nabla^3 \Delta\nu_{\rm Noise}$  & $ 1.32   _{- 118    }^{+ 133    } \times 10^{9}$ & $-3.3    _{- 1.06e+03}^{+ 944    } \times 10^{-14}$ & $-4.18   _{- 1.21e+04}^{+ 1.26e+04} \times 10^{-4}$ & $ 6.27   _{- 1.15e+03}^{+ 1.48e+03} \times 10^{-6}$ \\
\hline
$\nabla^4$ QPO norm               & $ 7.37   _{- 7.63   }^{+ 9.28   } \times 10^{3}$ & $ 1.46   _{- 1.5    }^{+ 1.49   } \times 10^{-23}$ & $-5.09   _{- 14.7   }^{+ 15.5   } \times 10^{-10}$ & $-2.02   _{- 2.99   }^{+ 3.81   } \times 10^{-12}$ \\
$\nabla^4\Delta\nu_{\rm QPO}$     & $ 8.32   _{- 9.24   }^{+ 10.9   } \times 10^{6}$ & $ 1.54   _{- 15.2   }^{+ 15.9   } \times 10^{-21}$ & $-5.21   _{- 139    }^{+ 158    } \times 10^{-8}$ & $ 8.48   _{- 44.1   }^{+ 52.9   } \times 10^{-10}$ \\
$\nabla^4\nu_{\rm QPO}$           & $-4.88   _{- 5.67   }^{+ 5.34   } \times 10^{6}$ & $ 3.2    _{- 9.16   }^{+ 9.98   } \times 10^{-21}$ & $ 2.23   _{- 10.3   }^{+ 9.92   } \times 10^{-7}$ & $-1.47   _{- 0.387  }^{+ 0.366  } \times 10^{-8}$ \\
$\nabla^4$ Noise norm             & $ 3.49   _{- 10.5   }^{+ 10.2   } \times 10^{3}$ & $ 1.22   _{- 1.68   }^{+ 1.67   } \times 10^{-23}$ & $ 1.32   _{- 2.11   }^{+ 2.06   } \times 10^{-9}$ & $ 2.91   _{- 2.98   }^{+ 2.98   } \times 10^{-12}$ \\
$\nabla^4 \Delta\nu_{\rm Noise}$  & $-3.07   _{- 742    }^{+ 726    } \times 10^{8}$ & $ 2.72   _{- 1.68e+03}^{+ 1.86e+03} \times 10^{-20}$ & $-3.73   _{- 1.22e+03}^{+ 1.05e+03} \times 10^{-5}$ & $-2.63   _{- 446    }^{+ 430    } \times 10^{-8}$ \\
\hline
$\nabla^5$ QPO norm               & $ 2.69   _{- 30     }^{+ 76.4   } \times 10^{4}$ & $ 3.65   _{- 6.51   }^{+ 6.59   } \times 10^{-29}$ & $ 9.91   _{- 5.58   }^{+ 5.73   } \times 10^{-11}$ & $ 7.49   _{- 46.6   }^{+ 43.7   } \times 10^{-15}$ \\
$\nabla^5\Delta\nu_{\rm QPO}$     & $ 8.64   _{- 8.79   }^{+ 10.3   } \times 10^{8}$ & $-2.12   _{- 6.46   }^{+ 7.29   } \times 10^{-26}$ & $ 7.24   _{- 5.22   }^{+ 5.02   } \times 10^{-8}$ & $ 1.28   _{- 0.771  }^{+ 0.756  } \times 10^{-10}$ \\
$\nabla^5\nu_{\rm QPO}$           & $ 3.82   _{- 6.33   }^{+ 7.11   } \times 10^{8}$ & $-8.1    _{- 38.9   }^{+ 42.4   } \times 10^{-27}$ & $ 7.19   _{- 33.6   }^{+ 33.3   } \times 10^{-9}$ & $ 6.52   _{- 6.75   }^{+ 6.64   } \times 10^{-11}$ \\
$\nabla^5$ Noise norm             & $-2.51   _{- 9.55   }^{+ 8.95   } \times 10^{5}$ & $ 5.64   _{- 7.19   }^{+ 7.08   } \times 10^{-29}$ & $-1.17   _{- 0.701  }^{+ 0.714  } \times 10^{-10}$ & $-2.53   _{- 3.26   }^{+ 3.47   } \times 10^{-14}$ \\
$\nabla^5 \Delta\nu_{\rm Noise}$  & $ 8.23   _{- 1.08e+03}^{+ 1.31e+03} \times 10^{8}$ & $ 7.8    _{- 1.84e+04}^{+ 1.89e+04} \times 10^{-27}$ & $ 5.71   _{- 2.57e+03}^{+ 2.75e+03} \times 10^{-7}$ & $ 8.59   _{- 2.34e+03}^{+ 2.5e+03} \times 10^{-11}$ \\
\hline
    \end{tabular}
    \caption{Table of QPO parameters for model including first five derivatives of each basic parameter relative to a property of each segment. $\nabla$ indicates the derivative wrt the value of the sorting property.}
    \label{tab:qpo_pars5}
\end{table*}

\begin{figure*}
    \centering
    \includegraphics[width=\textwidth]{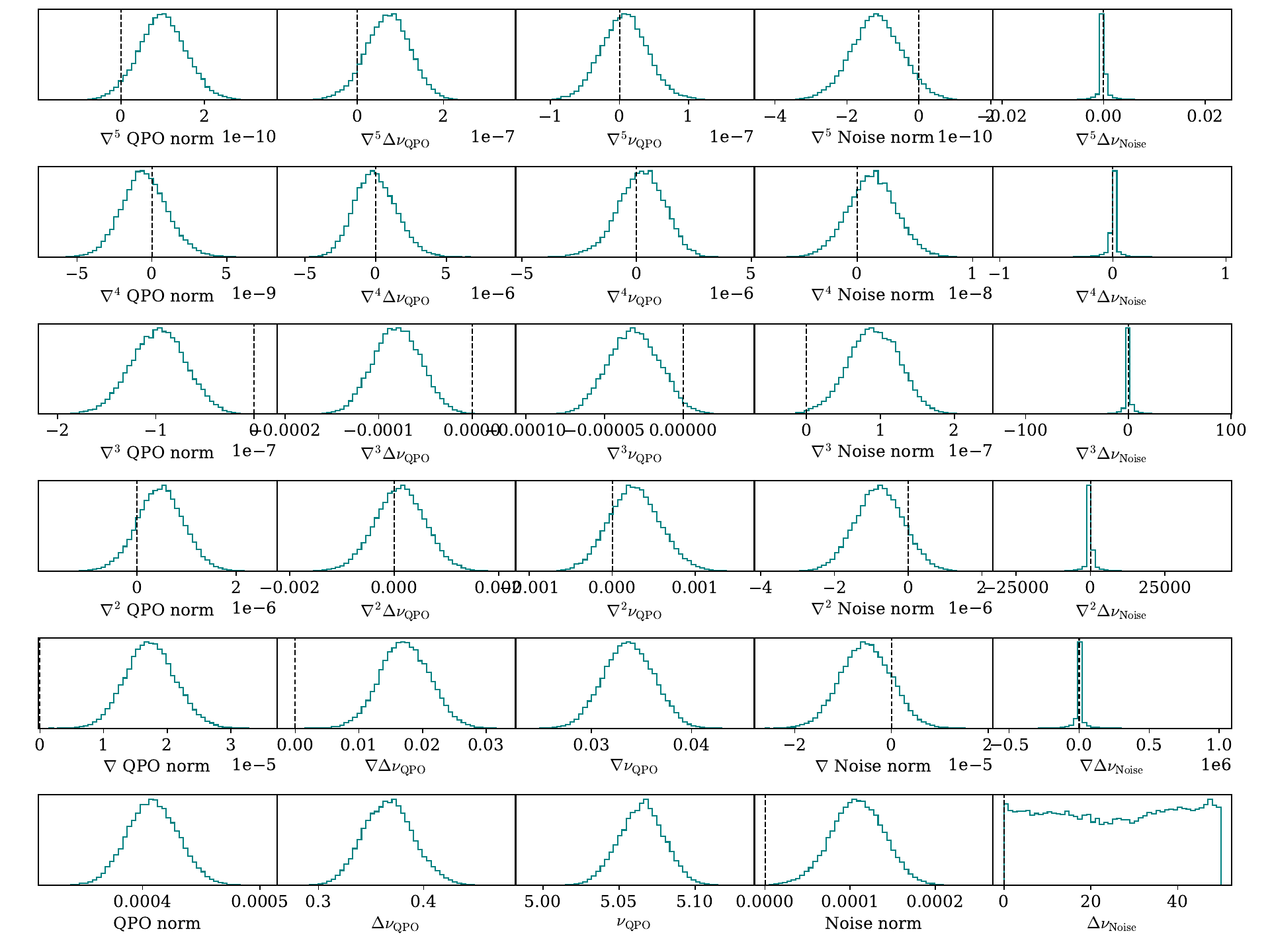}
    \caption{Parameter inferences (hard rate sorting; other sorting parameters are similar). All parameters except the noise width (and its derivatives) are unimodal with narrow tails; the noise width has little effect on the resulting model (see also Figure~\ref{fig:exampleCorner}). Zero is an acceptable value for the highest derivative of each parameter (this is the stopping criterion for the power series expansion).}
    \label{fig:examplePars}
\end{figure*}

\begin{figure}
    \centering
    \includegraphics[width=\textwidth]{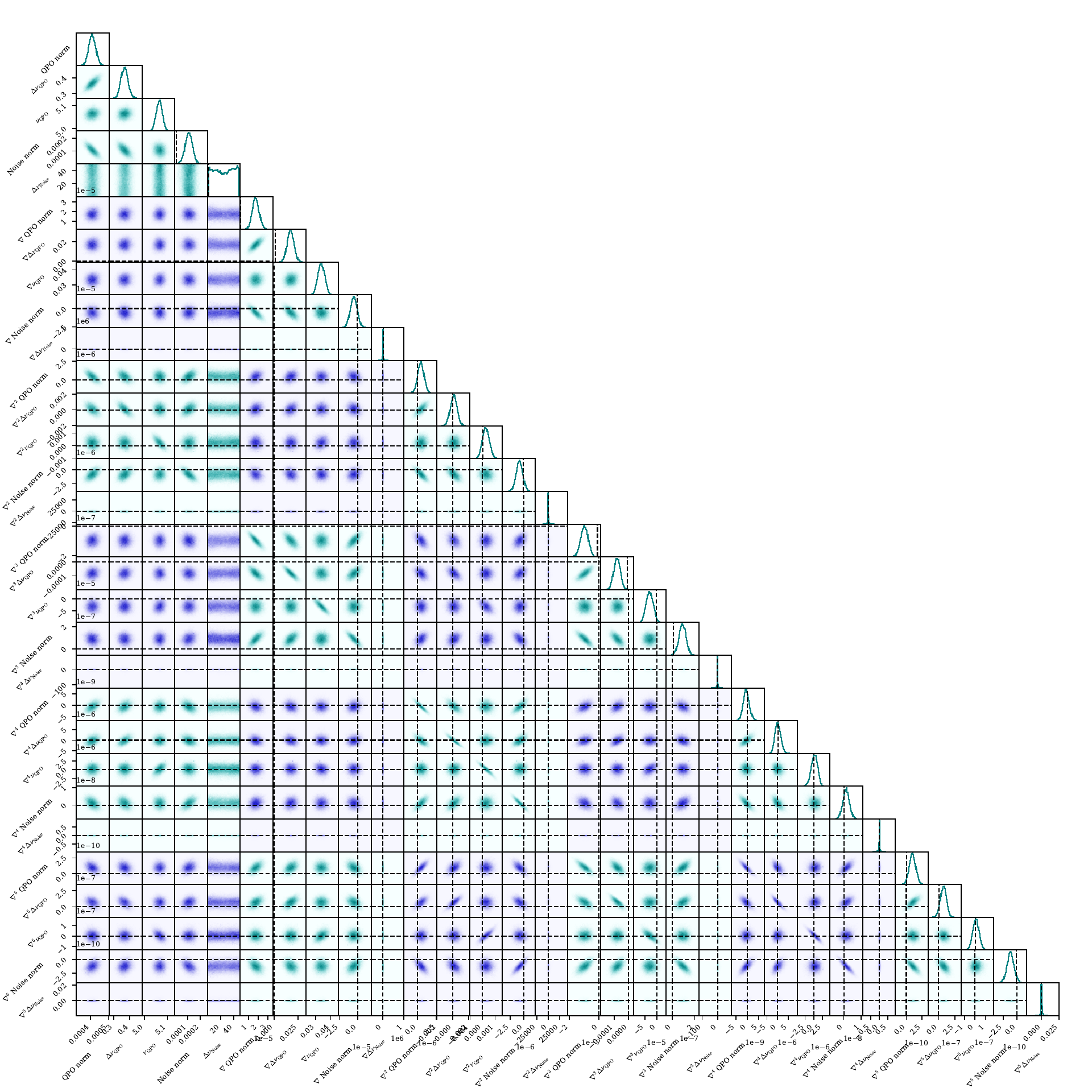}
    \caption{
    2D parameter inferences (hard rate sorting; other sorting parameters are similar). Parameters of interest do not correlate with unconstrained parameters.
    }
    \label{fig:exampleCorner}
\end{figure}

\begin{figure*}
    \centering
    \includegraphics[width=\textwidth]{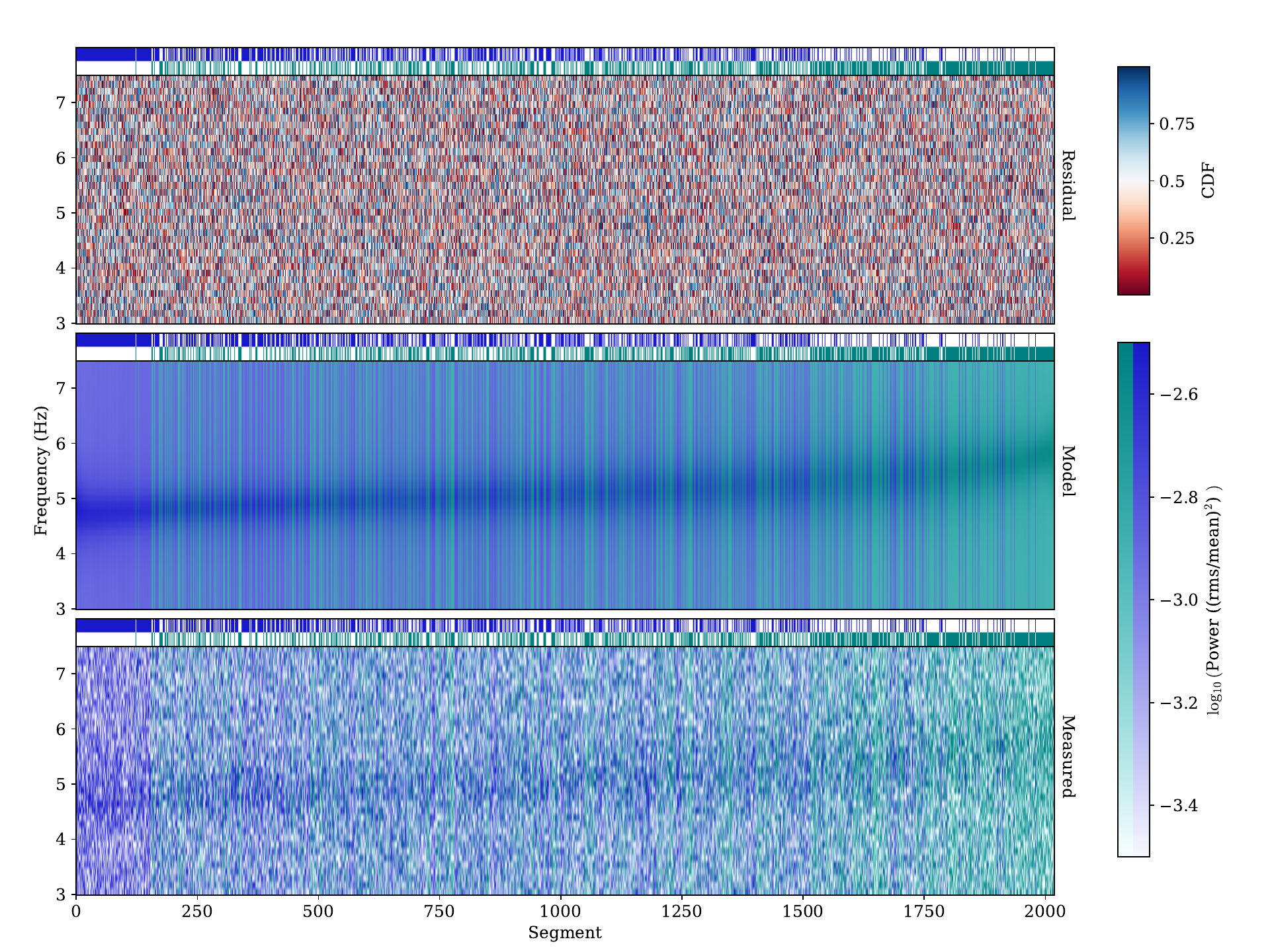}
    \caption{Model and residuals for maximum likelihood parameters (hard rate sorting)
    {\it Bottom:} Measured periodogram values;
    {\it Middle:} Model power spectrum with evolution as a function of hard rate;
    {\it Top:} Fractional residuals, as the quantile of the model CDF.
    }
    \label{fig:exampleModel}
\end{figure*}

\begin{figure*}
    \centering
    \includegraphics[width=\textwidth]{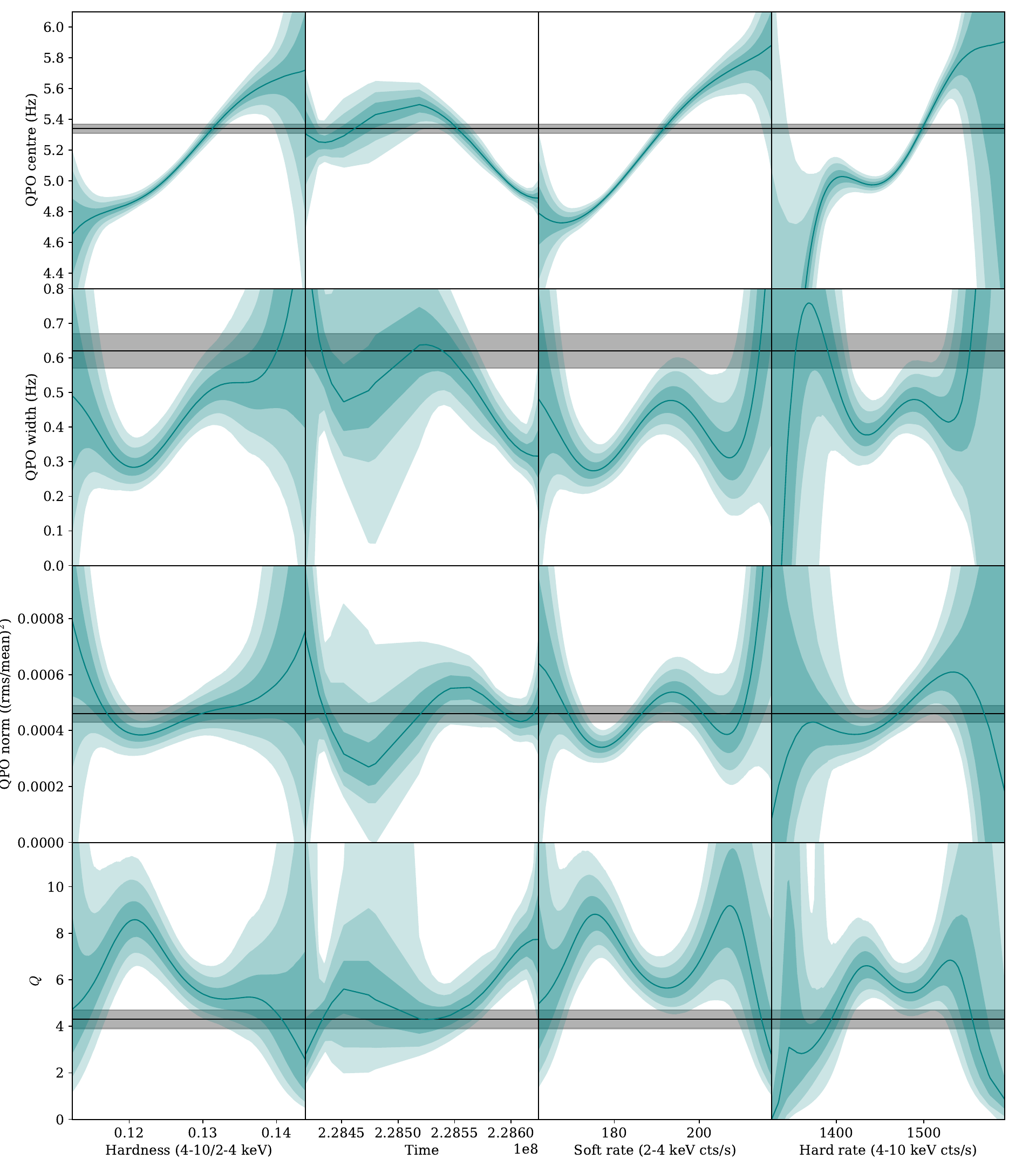}
    \caption{QPO parameters of fit (see text for details) to the dynamic power spectrum (blue) compared to the mean power spectrum over the same interval (black).
    The four columns show different values by which the segments in the dynamic power spectrum are sorted.
    The dynamic fit captures the evolution of the QPO frequency; the width is narrower as the smearing due to the frequency change is removed; and the quality factor is correspondingly higher.}    
    \label{fig:psdFitPars}
\end{figure*}

\begin{figure}
    \centering
    \includegraphics[width=\columnwidth]{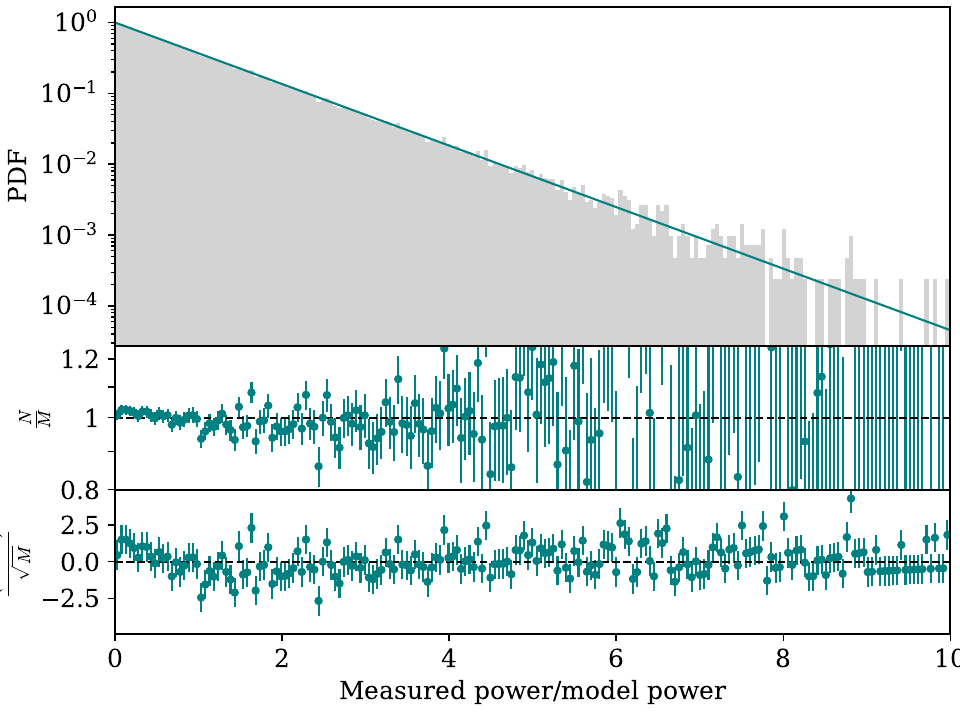}
    \includegraphics[width=\columnwidth]{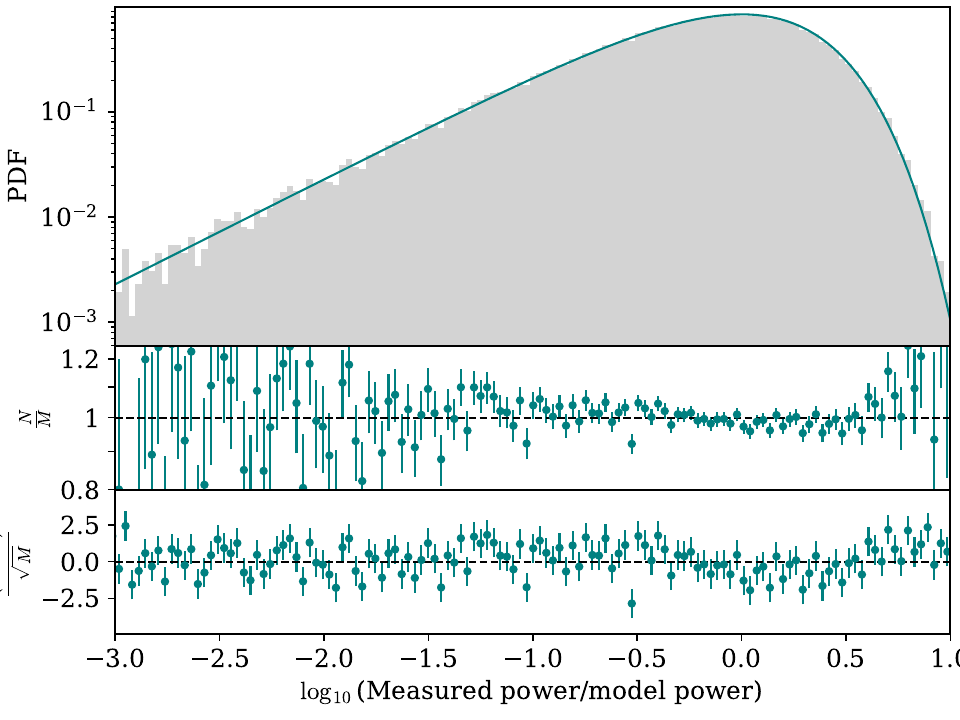}
    \caption{Comparison of empirical to proposed distribution of power spectral values, on both a linear ({\it upper set}) and logarithmic ({\it lower set}) scale.
    {\it Per set:}
    {\it Top:} PDF of a $\chi_2^2$ distribution (green line) and of the measured values (histogram);
    {\it Middle:} ratio of observed to expected values;
    {\it Bottom}: difference normalised by $\sqrt{M}$ (distributed approximately as a standard normal for large $N$).
    }
    \label{fig:distTest}
\end{figure}

\section{Additional figures}

\begin{figure*}
    \centering
    \includegraphics[width=0.85\textwidth]{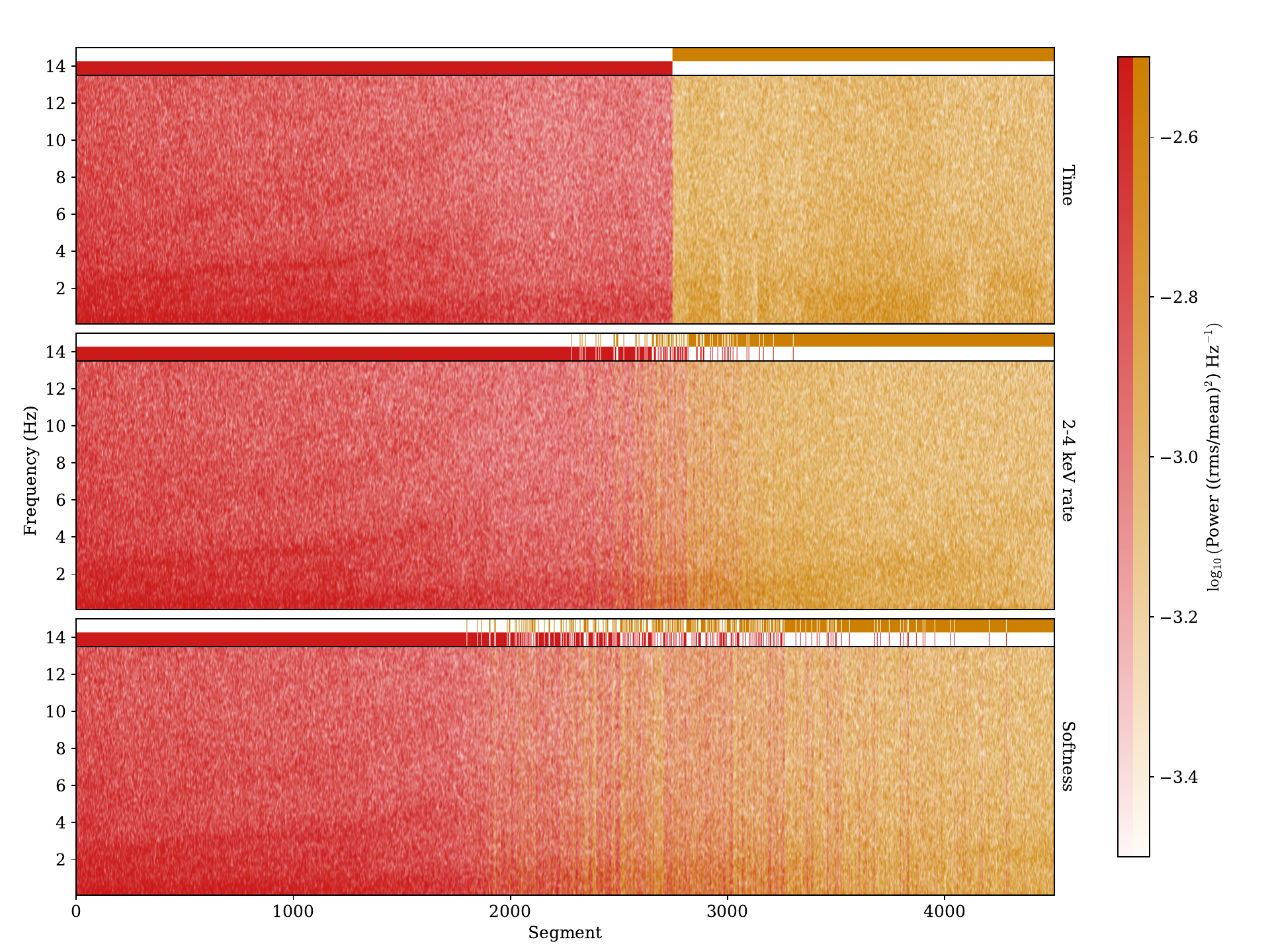}
    \includegraphics[width=0.85\textwidth]{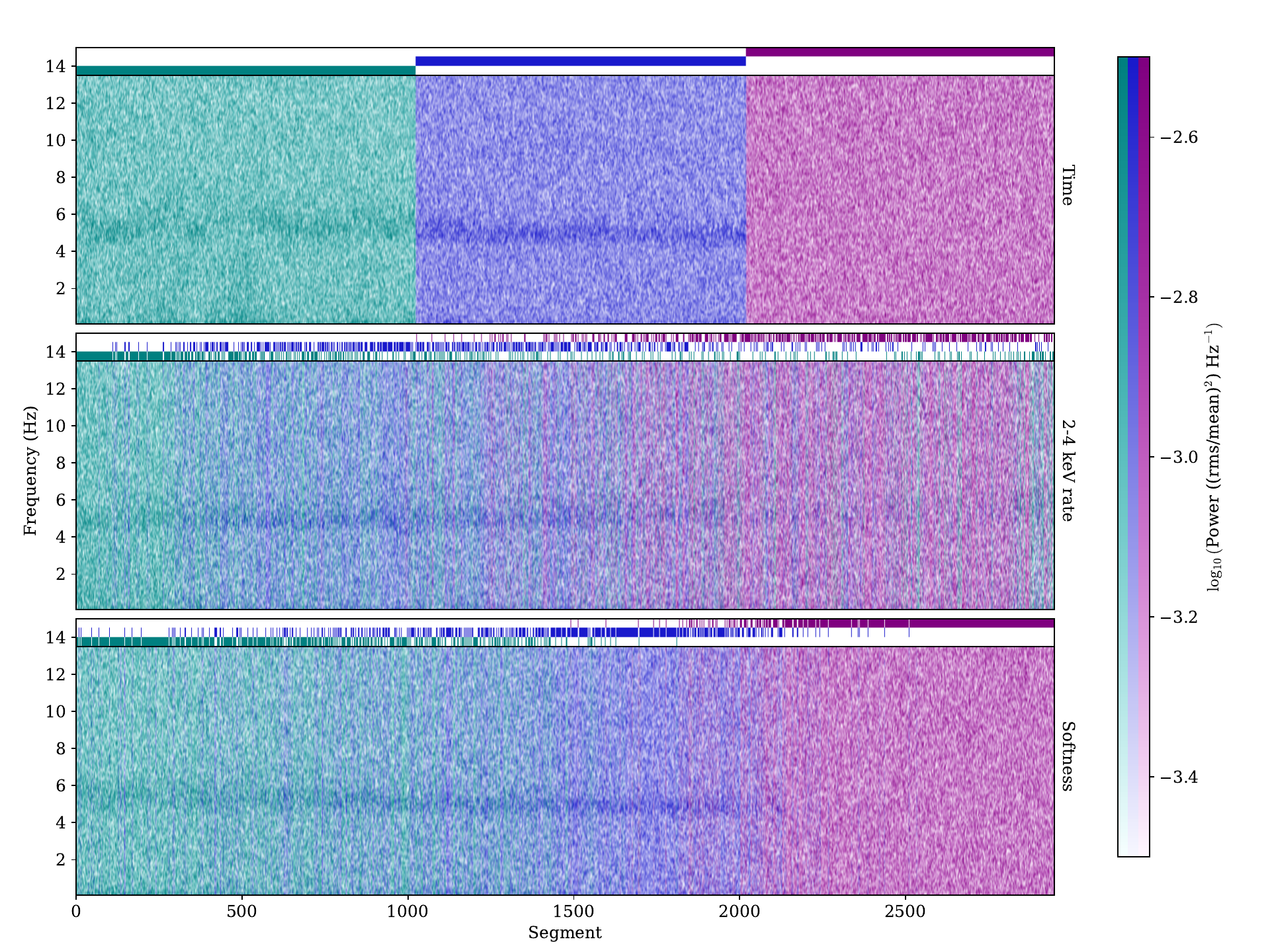}
    \caption{Version of Figure~\ref{fig:dynamic_psds} with hard and soft sides separately: dynamical power-spectrum of the observations chosen in this paper. Top: sorted by time, with time gaps removed. Middle: sorted by observed soft ($2-4$ keV) count rate. Bottom: sorted by softness, i.e. ratio of soft ($2-4$ keV) to hard ($4-10$ keV) count rates. On all panels, segments are $8$ seconds long and the different states are shown in their associated colours (Figure~\ref{fig:hid}).}
    \label{fig:dpsd_separate}
\end{figure*}

\begin{figure*}
    \centering
    \includegraphics[width=\textwidth]{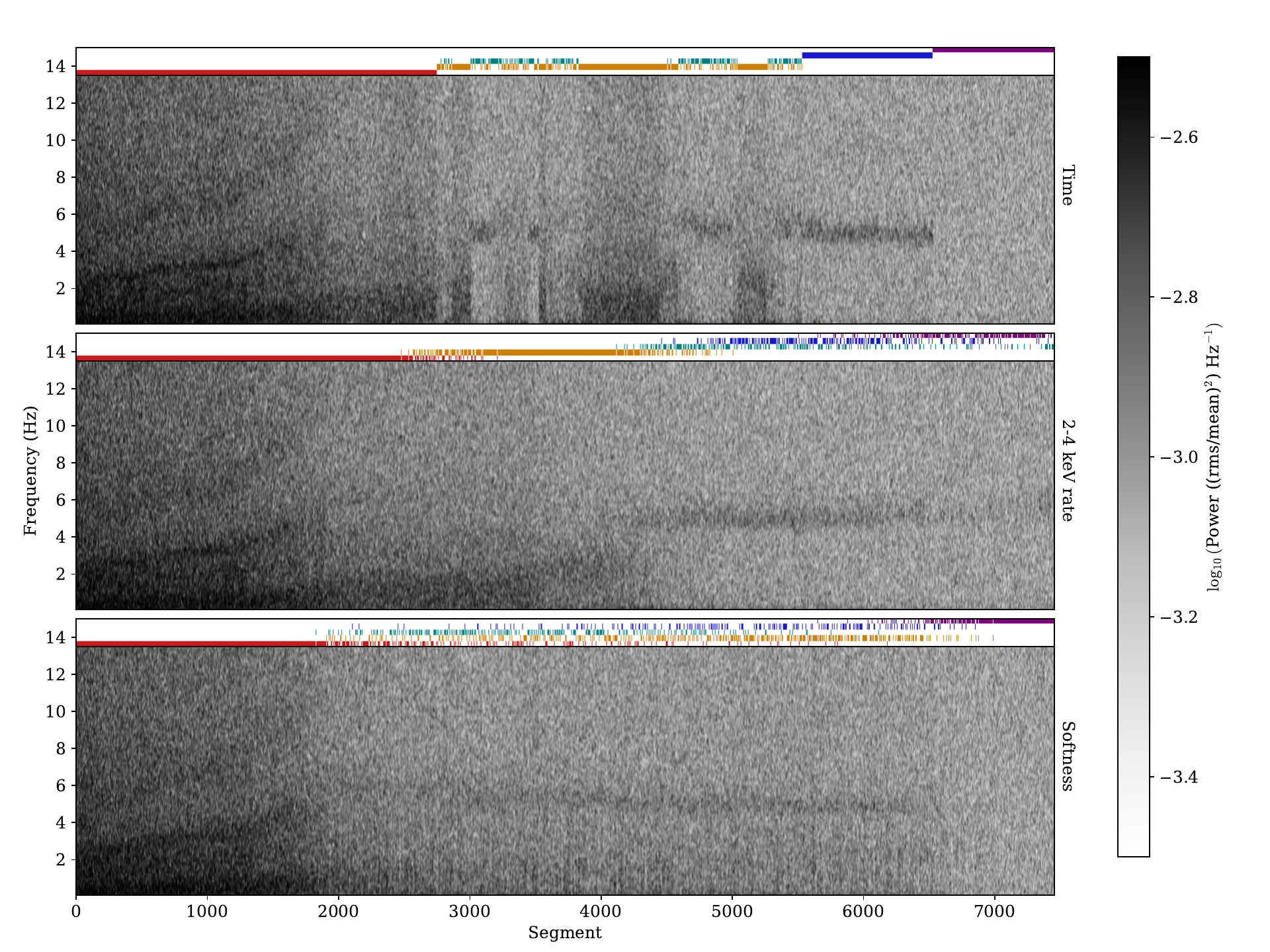}
    \caption{Greyscale version of Figure~\ref{fig:dynamic_psds}: dynamical power-spectrum of the observations chosen in this paper. Top: sorted by time, with time gaps removed. Middle: sorted by observed soft ($2-4$ keV) count rate. Bottom: sorted by softness, i.e. ratio of soft ($2-4$ keV) to hard ($4-10$ keV) count rates. On all panels, segments are $8$ seconds long and the different states are shown in their associated colours (Figure~\ref{fig:hid}).}
    \label{fig:dynamic_psds_grey}
\end{figure*}

\bsp
\label{lastpage}
\end{document}